\documentclass[journal,twoside,web]{ieeecolor}
\usepackage{generic}
\usepackage{amsmath,amssymb,amsfonts}
\usepackage{algorithmic}
\usepackage{graphicx}
\usepackage{textcomp}
\usepackage{booktabs}
\usepackage{bbold}

\DeclareMathAlphabet{\mathpzc}{OT1}{pzc}{m}{it}
\usepackage{bm}
\usepackage{cite}
\usepackage{caption}
\usepackage{subcaption}
\usepackage{optidef}
\usepackage[bookmarks=false,draft]{hyperref}
\def\BibTeX{{\rm B\kern-.05em{\sc i\kern-.025em b}\kern-.08em
    T\kern-.1667em\lower.7ex\hbox{E}\kern-.125emX}}
\begin{document}
\title{SleepPoseNet: Multi-View Learning for Sleep Postural Transition Recognition \\Using UWB}

\author{Maytus~Piriyajitakonkij, Patchanon~Warin, Payongkit~Lakhan, Pitsharponrn~Leelaarporn, Nakorn~Kumchaiseemak, Supasorn~Suwajanakorn, Theerasarn~Pianpanit, Nattee~Niparnan, Subhas~Chandra~Mukhopadhyay~\IEEEmembership{Fellow, IEEE} and Theerawit~Wilaiprasitporn~\IEEEmembership{Member, IEEE}
\thanks{This work was supported by PTT Public Company Limited, The SCB Public Company Limited, Thailand Science Research and Innovation (SRI62W1501) and Office of National Higher Education Science Research and Innovation Policy
Council (C10F630057).}
\thanks{M. Piriyajitakonkij, P. Lakhan, P. Leelaarporn and T. Wilaiprasitporn are parts of Bio-inspired Robotics and Neural Engineering Lab, School of Information Science and Technology, Vidyasirimedhi Institute of Science \& Technology, Rayong, Thailand {\tt\small (corresponding authors: theerawit.w at vistec.ac.th)}.}
\thanks{N. Kumchaiseemak and S. Suwajanakorn are with Vision and Learning Lab, School of Information Science and Technology, Vidyasirimedhi institute of Science \& Technology, Rayong, Thailand}
\thanks{T. Pianpanit is with Department of Applied Radiation and Isotopes, Faculty of Science, Kasetsart University, Bangkok, Thailand}
\thanks{P. Warin and N. Niparnan are with the Computer Engineering Department, Chulalongkorn University, Bangkok, Thailand.}
\thanks{S. C. Mukhopadhyay is with School of Engineering, Macquarie University, Sydney, NSW 2109, Australia}
\thanks{Raw dataset, code examples, and other supporting materials are available on \url{https://github.com/IoBT-VISTEC/SleepPoseNet} (available when this manuscript is accepted.).}}

%
%

\markboth{Journal of \LaTeX\ Class Files,~Vol.~14, No.~8, August~2015}%
{Shell \MakeLowercase{\textit{et al.}}: Bare Demo of IEEEtran.cls for IEEE Journals}
%



\maketitle

\begin{abstract}

\textcolor{blue}{Recognizing movements during sleep is crucial for the monitoring of patients with sleep disorders, and the utilization of ultra-wideband (UWB) radar for the classification of human sleep postures has not been explored widely. This study investigates the performance of an off-the-shelf single antenna UWB in a novel application of sleep postural transition (SPT) recognition. The proposed Multi-View Learning, entitled SleepPoseNet or SPN, with time series data augmentation aims to classify four standard SPTs. SPN exhibits an ability to capture both time and frequency features, including the movement and direction of sleeping positions. The data recorded from 38 volunteers displayed that SPN with a mean accuracy of $73.7 \pm 0.8 \%$ significantly outperformed the mean accuracy of $59.9 \pm 0.7 \%$ obtained from deep convolution neural network (DCNN) in recent state-of-the-art work on human activity recognition using UWB. Apart from UWB system, SPN with the data augmentation can ultimately be adopted to learn and classify time series data in various applications.}

\end{abstract}

\begin{IEEEkeywords}
sleep pose recognition, multi-view deep learning, ultra-wideband (UWB), deep canonical correlation analysis (DCCA), deep convolution neural network (DCCN), time series data augmentation.
\end{IEEEkeywords}

%
\IEEEpeerreviewmaketitle
\section{Introduction}
\IEEEPARstart{S}{leep} 
is one of the most essential elements of human health. Sleep disorders such as obstructive sleep apnea (OSA), are associated with cardiovascular disease (CVD), stroke, hypertension, and daytime sleepiness \cite{drager2017sleep, mcdermott2020sleep, bixler2019association}, resulting in the lack of work productivity, car accidents, and a higher rate of mortality \cite{jurado2015workplace, garbarino2016risk}.
One important aspect with definite advantages to determine the quality of sleep and the severity of the disorders is to monitor bodily spatial movement, especially different postures during the night \cite {walsh2016noncontact, lin2016sleepsense, rundo2019polysomnography}. OSA severity can be identified by using sleep postures as one of the important variables.  Furthermore, numerous individuals at risk who are experiencing pressure ulcers or bedsores, which are lesions on the skin resulting from pressure, may face a higher mortality rate, especially when they are bedridden with the chronic type \cite{lyder2003pressure, yousefi2011bed}. Thus, sleep monitoring can be used to notify the caregivers in order to adjust the sleep postures of the patients, increasing the possibility of pressure ulcers prevention.

Technological advances in the Internet of Things (IoT) and deep learning have been used to unleash many in-home sleep monitoring devices through the consumer markets \cite{Fallmann}. These devices consist of two types: wearable and non-wearable. Both types have been utilized to recognize the physical position and monitor the movement of patients with sleep disorders \cite{Cao}. Most of the wearable sleep monitoring devices are developed as smartwatches or smartbands for simplicity and equipped with features to track crucial physiological features, such as heart rate (HR) and oxygen saturation, using Photoplethysmography (PPG) \cite{li2017evaluation}. Some devices are also equipped with accelerometer and gyroscope for human activity recognition and sleep stages estimation \cite {sun2017sleepmonitor}. Although wearable devices have been proved to provide accurate HR estimation \cite{tanut}, the measurement can only take place on the location where the device is worn such as the upper limb. This may result in high false positive rates and may be inconvenient. Therefore, using non-wearable devices to track body movement is more preferable. Different non-wearable devices have been adopted for sleep monitoring including camera \cite{deng2018design}, pressure mat \cite{yousefi2011bed, liu2013dense, xu2016body}, and radar-based system \cite{walsh2016noncontact}. Despite the advantages offered by these devices, some hindrances can emerge: Cameras are generally susceptible to different light conditions \cite{deng2018design} and privacy issue may be concerned when using an infrared camera. Pressure mat is needed to place appropriately as the patients’ movements may shift its placement and require re-positioning. In contrast, a radar-based sensor does not face the same dilemmas and has high penetration ability which can detect even through-wall human movements \cite{li2012through} and vital signs \cite{wang2018through}.



Many studies investigating radar-based systems for sleep monitoring applications have been exploring their ability to detect human vital signs, such as HR and breathing rate \cite{ shyu2018detection, shen2018respiration, lee2018novel, biocas1, subhas2}. These systems can also be used to identify breathing disorders \cite{8322142}, body movements and HR in different sleep stages \cite{rahman2015dopplesleep, hong2018noncontact}. For instance, Kim \textit{et al.} applied Continuous Wave (CW) Doppler radar and Deep Convolutional Neural Networks (DCNN), which exploited knowledge in Doppler effect of reflected signals to obtain the velocity of targets, to detect human and classify activity, such as hand gestures \cite{kim2016hand}, with higher accuracy \cite{kim2015human}. However, a definite drawback of CW radar is presumably due to the lack of range information in the signals. To solve this issue, a Frequency Modulated Continuous Wave (FMCW) radar, of which the signals contain both range and Doppler information, was used as an alternative \cite{dingFMCW}. However, multipath interference seems to reduce the performance of FMCW radar, especially for an indoor application.


Another attractive spectrum, the Ultra-Wideband (UWB) radar, has been utilized for its range information with high resolution due to its high frequency pulse signals \cite{fontana2004recent, zhang2006accurate}. UWB has been proposed to be more appropriate to an indoor usage due to a higher multipath suppression capability with more affordable price \cite{zhang2006accurate}. In 2016, Yin \textit{et al.} introduced a combination of ECG and UWB radar in order to classify cardiac arrhythmia and found that the accuracy had improved from using only ECG \cite{yin2016ecg}. To evaluate the classification of human activity recognition, many studies used classical machine learning methods such as Support Vector Machine (SVM) \cite{kim2009human, javier2014application}. Bryan \textit{et al.} used  Principal Component Analysis (PCA) to reduce dimensions and extract features in time domain combining with the frequency components extracted by Fourier Transform (FT) \cite{bryan2012application}. Ding \textit{et al.} created new method for UWB feature extraction called Weighted Range-Time-Frequency Transform (WRTFT), which included range information to a Short Time Fourier Transform (STFT) \cite{ding2018non}. WRTFT was later used by Chen \textit{et al.} with DCNN, obtaining more robust result \cite{chen2019non}.

With the promising results of the application of the radar-based system on human activity classification tasks, some studies have turned its interest towards analyzing on-bed motions \cite{rahman2015dopplesleep, lin2016sleepsense, nguyen2016detection, lee2016movement}. However, despite the extensive usages of UWB for physical activity recognition, to our best knowledge, none has examined its utilization for detecting and classifying sleep postures. In this study, the development of a deep learning algorithm in order to make an accurate classification of Sleep Postural Transitions (SPT) was investigated. Off-the-shelf UWB radar device was employed in a smart bedroom environment. Due to the ambiguity of the data between classes, time domain features were fused with frequency domain features by Multi-View Learning (MVL) \cite{review_mtv} to improve classification accuracy. Our method was then compared to the state-of-the-art method to determine the most optimal method for SPT recognition.

Three main contributions are presented in this study:

\begin{enumerate}
{\color {blue}
\item To our best knowledge, this is the first study to propose using Sleep Postural Transitions (SPT) classification models through off-the-shelf Ultra-Wideband (UWB) radar system. The model exhibits the potential to support the usage of the current sleep monitoring systems on measurement of vital signs and sleep stages. Moreover, the effect of an unrelated moving object in a bedroom-like environment on UWB classification performance was examined. Different participant distributions in training data and testing data were also investigated.}
{\color {blue}
\item Multi-View Learning (MVL) model was used to fuse information in time and frequency domains, introducing a new method for UWB signals recognition, entitled SleepPoseNet or SPN. 
}
\item Various data augmentation techniques for time series were investigated and employed in an attempt to increase the classification accuracy.

\end{enumerate}

\section{Methods}
\subsection{UWB Principle and Hardware}
UWB radar propagates the impulse signals through a transmitter \cite{cassioli2002ultra}. Once the pulse arrives at the focused object, the pulse is split into two; the reflected pulse returns to the signal receiver, while the transmitted pulse passes through the object. Time of Arrival (TOA) is measured to identify the range of the object. A received signal at an antenna is represented by \autoref{eq:1};
\begin{equation}\label{eq:1}
y_{k}(t) = \sum_{i=1}^{N_{path}}a_{ki}x(t-\tau_{ki})+n(t)
\end{equation}

where $y_{k}(t)$ is the received signal of the $k$-th pulse sending from the transmitter and $k$ is called a slow time index. $x$ and $a_{ki}$ represent the original signals sent from transmitter and the scaling value of each reflected signal. $i$ means $i$-th path of travelling pulse between the transmitter and receiver, and $N_{path}$ is the number of the travelling paths. $t$ is the fast time index, representing the arrival time of each reflected signal. $\tau_{ki}$ is the time delay, and $n(t)$ is noise. The range $R$ between device and object can be calculated by the TOA $\Delta t$ and the speed of light $c$ as in an \autoref{eq:2};
\begin{equation}\label{eq:2}
R = \frac{c\Delta t}{2}
\end{equation}
\

The data were collected using the Xethru X4M03 development kit, a state-of-the-art UWB radar device invented by Novelda \cite{X4m03}. Its transceiver operates within the range of 5.9-10.3 GHz. The key features include its flexibility to be customized and suitable for particular application, i.e., proper configurations can be primed to support our applications, such as detection range, number of frames per second, number of pulses per step, and number of sweep iterations. Moreover, the device supports digital down conversion, filtering, and data rate reduction, which are efficient signal processing methods. The radar configurations in our experiments are shown in \autoref{table: Radar Prams} and the detailed information are described in \cite{X4}.

\begin{table}
\caption{Radar Parameters.}
\centering 
\begin{tabular}{p{4cm} p{4cm}} 
\toprule 
\textbf{Parameters} & \textbf{Values} \\
\midrule 
Tx  & 7.29 GHz (center frequency) \\
Pulse Repetition Frequency  &  15.18 MHz \\
Sampling Frequency  & 23.32 GHz\\

Range &  0-9 m \\

Bins per radar frame &  180 (baseband)\\
Frame rate  &  10 \\
\bottomrule
\end{tabular}
\label{table: Radar Prams}
\end{table}

\subsection{Experimental Recording}
{\color {blue}
The UWB radar system was attached to the wall with 0.8 m height above the bed. The pitch angle was approximately 45 degrees downward towards the bed as shown in \autoref{fig:bedroom}. 
Two datasets comprised the recordings from 26 and 12 volunteers. Dataset I contained only 4 SPTs, and Dataset II held 5 classes, including 4 SPTs and background (BG). In addition, Dataset II consisted of two sessions: without and with a moving object.
}
\subsubsection{Dataset I}

\begin{figure}[]
     \centering
     \begin{subfigure}[]{0.1\textwidth}
         \centering
         \includegraphics[width=\textwidth]{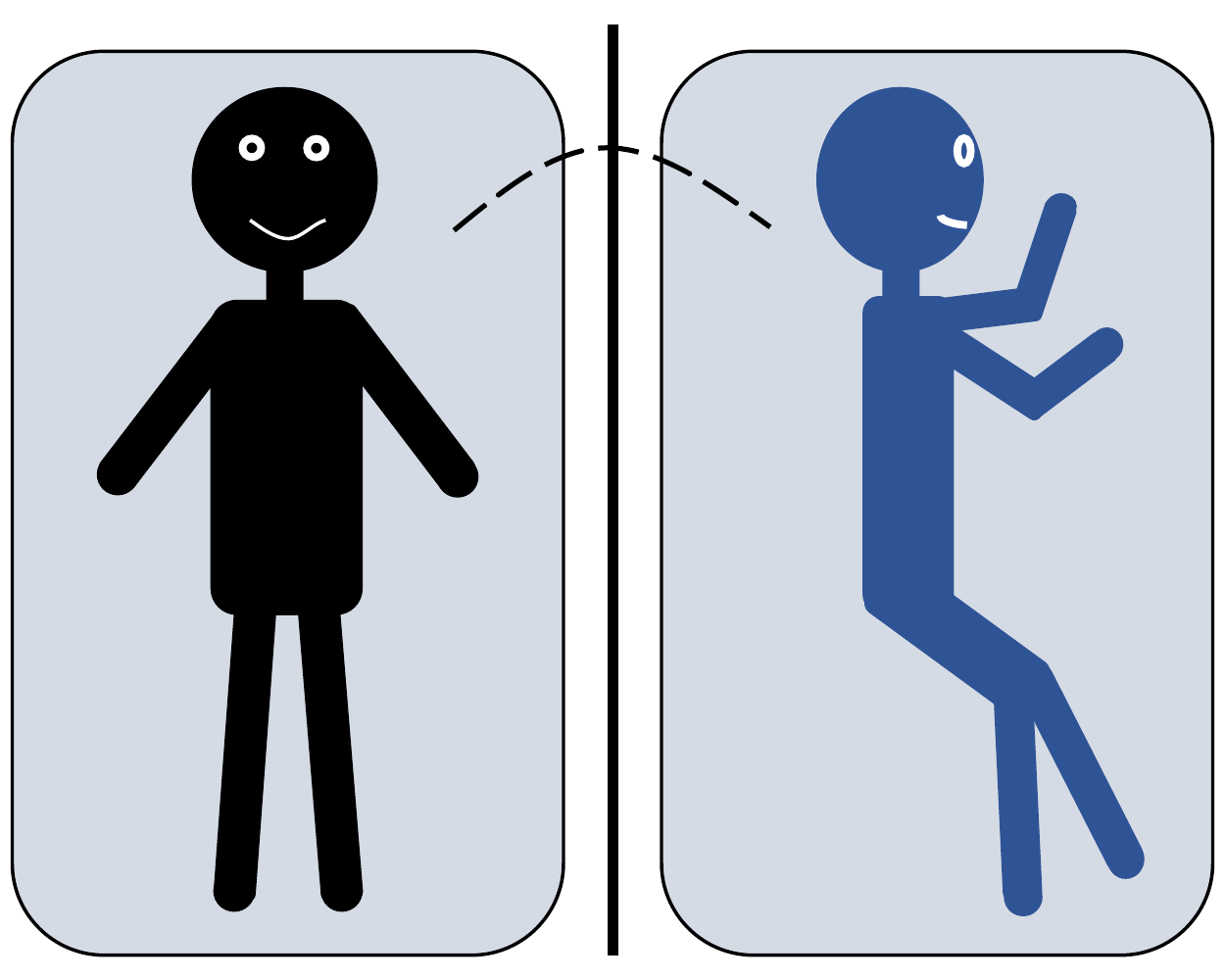}
         \caption{}
         \label{fig:post_susi}
     \end{subfigure}
     \hfill
     \begin{subfigure}[]{0.1\textwidth}
         \centering
         \includegraphics[width=\textwidth]{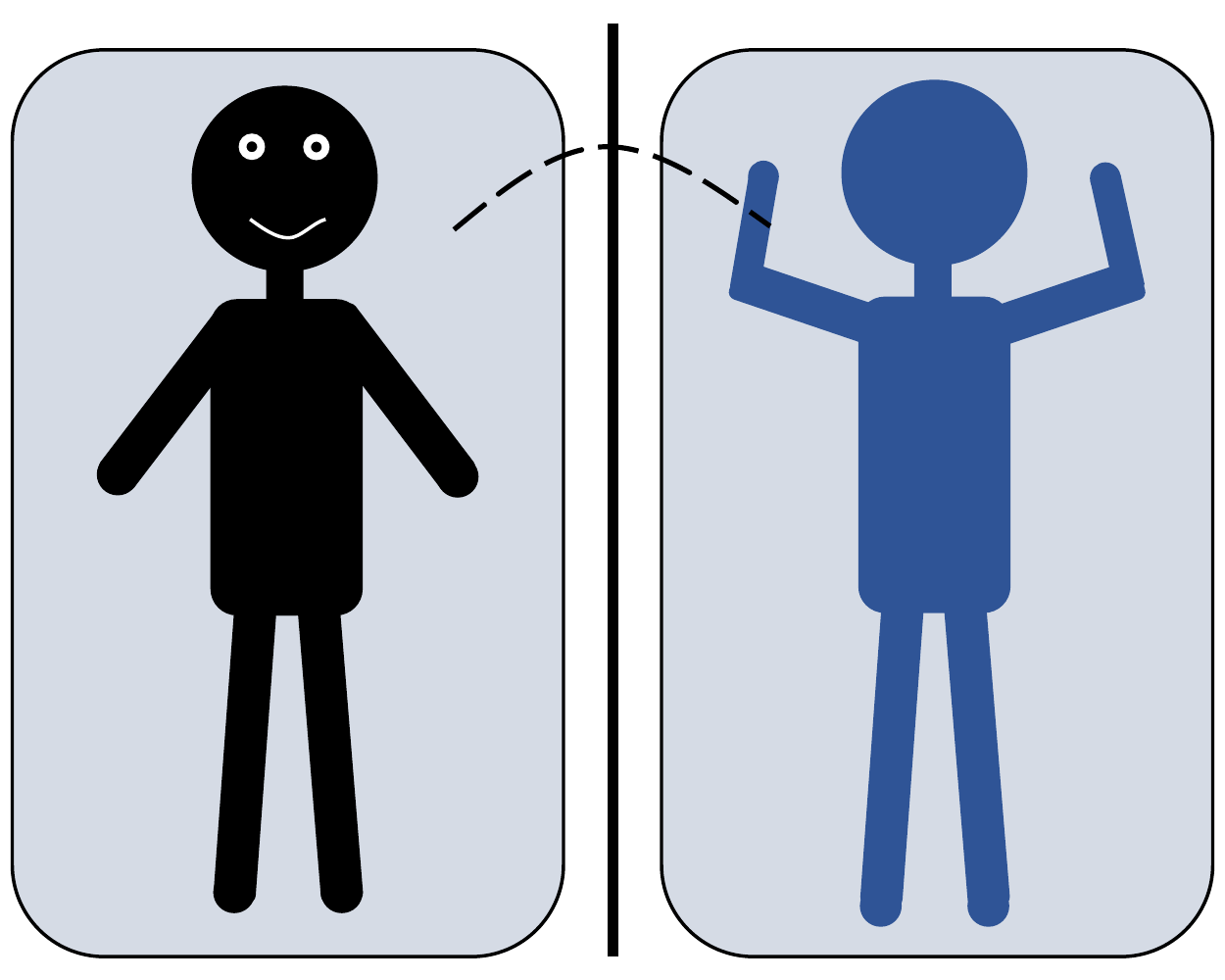}
         \caption{}
         \label{fig:supr}
     \end{subfigure}
     \hfill
     \begin{subfigure}[]{0.1\textwidth}
         \centering
         \includegraphics[width=\textwidth]{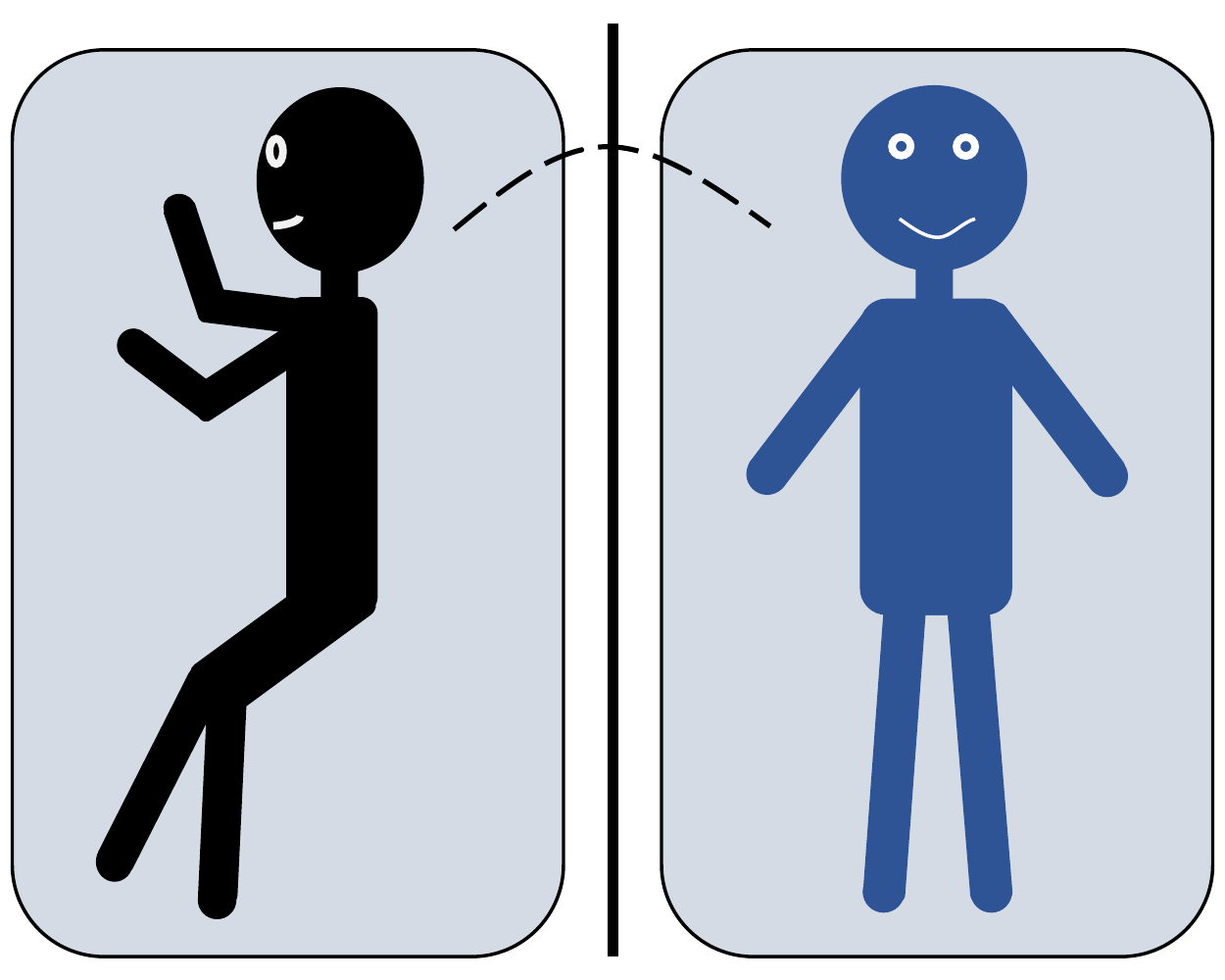}
         \caption{}
         \label{fig:sisu}
     \end{subfigure}
     \hfill
     \begin{subfigure}[]{0.1\textwidth}
         \centering
         \includegraphics[width=\textwidth]{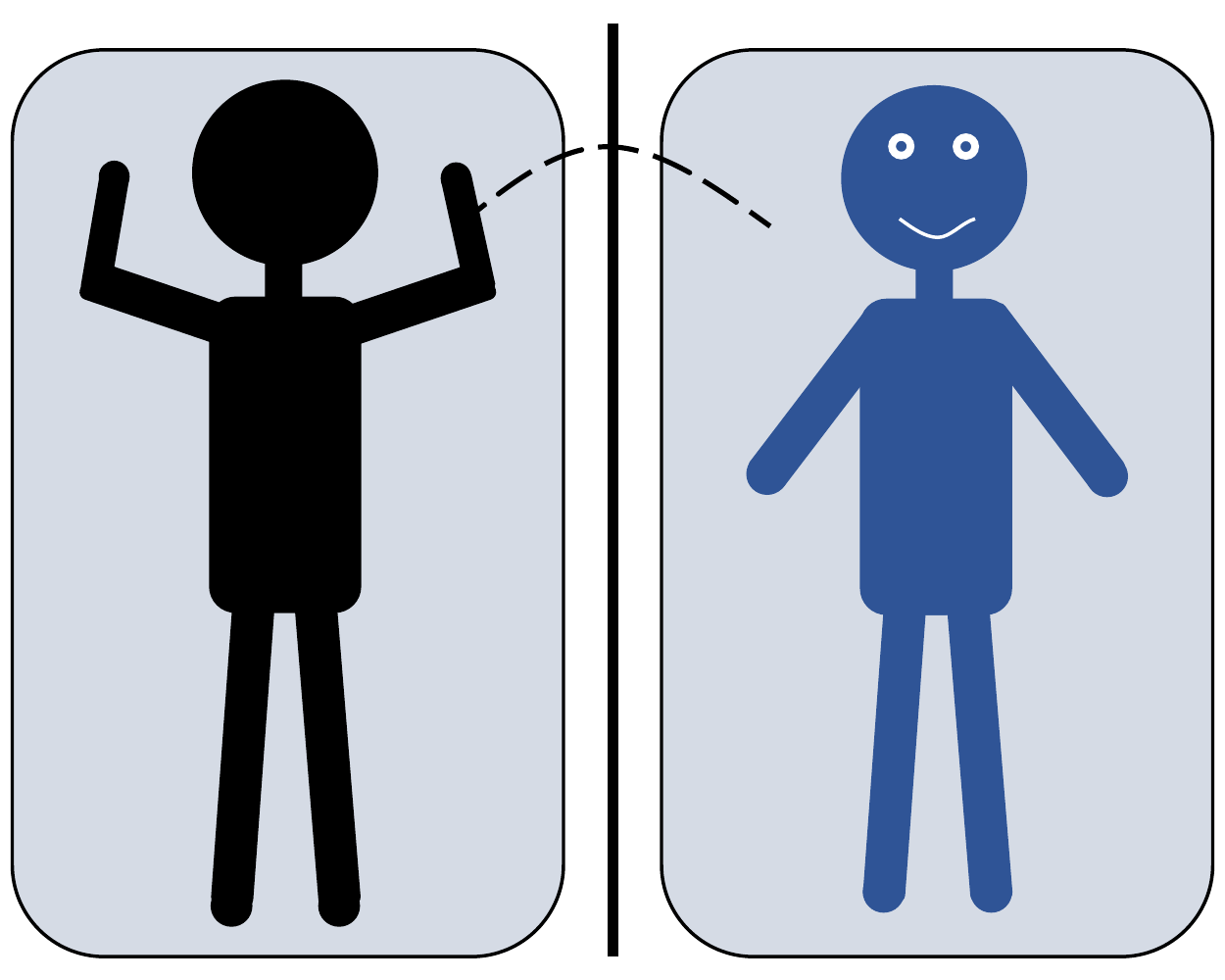}
         \caption{}
         \label{fig:prsu}
     \end{subfigure}     
        \caption{Sleep Postural Transitions (SPTs). (a) Supine to Side (SUSI) (b) Supine to Prone (SUPR) (c) Side to Supine (SISU) (d) Prone-to-Supine (PRSU)}
        \label{fig:sleep_post}
\end{figure}

A total of 26 volunteers, including 19 males and 7 females, participated in this study. The height ranged from 1.55 to 1.80 m, the weight ranged from 40 to 90 kg, and the age ranged from 18 to 35 years old. The experiment commenced with the participants lying down with a supine position in the middle of the bed under the line-of-sight of the radar. The participants were instructed to perform six motions, in a sample, following an arranged sequence: supine to left lateral, left lateral to supine, supine to right lateral, right lateral to supine, supine to prone, and prone to supine, at their own pace. The participants were given 10-15 s to rest while inhaling for 4-5 times before performing the next motion. This allowed the radar system to obtain the respiration features before and after each change of posture. Most participants performed one experimental set, consisting of 30 samples with five samples per class, whereas three participants performed two sets and one participant performed three sets. Furthermore, left and right lateral positions were treated as side sleep positions. Ultimately, a total of 31 experimental sets, consisting of four classes of SPT, designated as Supine to Side (SUSI), Supine to Prone (SUPR), Side to Supine (SISU), and Prone to Supine (PRSU), were recorded as shown in \autoref{fig:sleep_post} (a-d). To attain balanced classes, some of the samples of the positions SUSI and SISU in every set were undersampled. Therefore, a total of 620 samples for all experiments were finalized.

{\color {blue}
\subsubsection{Dataset II} 
There are 12 volunteers, including 9 males and 3 females participated in this study. The height ranged from 1.55 to 1.80 m, the weight ranged from 40 to 90 kg, and the age ranged from 20 to 30 years old. Dataset II was divided into two sessions. Session I contained the recordings of the participants in an environment in which all objects remained motionless. Whereas in Session II, a swinging fan was installed, representing a more realistic setting. The experimental protocols in both sessions were identical. Each participant performed 4 SPTs, similar to Dataset I, for a total of 6 times. In order to simulate naturalistic scenarios, each participant was instructed to perform other activities, including moving the limbs, head, and body, using smartphone, and lying motionlessly, without changing the sleep position. These activities were labeled as Background (BG). 10 background samples per participant in each session were recorded. To simulate the uncertainty of different users, the pitch angle of UWB radar was slightly adjusted. The position of the moving fan in the bedroom was changed for every participant. Hence, a total of 816 samples from all participants were recorded.
}

The radar signals from each sample was labelled manually using the video recordings from 10-15 minutes experiment as the ground truth. Each sample contained 10 s length where the first 3-4 s began before the first posture change. Informed consents were received from all participants following the Helsinki Declaration of 1975 (as revised in 2000), which was approved by the internal review board of Rayong Hospital, Thailand (RYH REC No.E010/2562).
\subsection{Data Processing}
After sampling, obtained signals are stored in $M \times N$ matrix $R$, where $M = 180$ is the number of fast time indices, also called range bins, and $N = 160$ is the number of slow time indices. $R_{mn}$ is the entry in $m$-th row and $n$-th column of matrix $R$. Subsequently, the following two steps are applied before extracting the meaningful features.
\paragraph{DC Suppression} The data in each fast time index or range bin $n$ may unavoidably contain DC noise which can be eliminated by averaging its value through all the number of slow time indices $N$.

\begin{equation}\label{eq:time_avg}
\bar{R}_{mn} = R_{mn} - \frac{1}{N}\sum_{i=0}^{N-1}R_{mi}
\end{equation}
The second term is an average by column. Therefore, every $R_{mn}$ is subtracted by its mean along the slow time index, before storing in $\bar{R}_{mn}$.
\paragraph{Background Suppression} The information at an instant slow time composed of static clutter, which are the uninterested objects and obstacles in an environment, and interested target, which referred to an active human. To suppress the static clutter, the data in each slow time index are subtracted with their average along fast time.

\begin{equation}\label{eq:range_avg}
Y_{mn} = \bar{R}_{mn} - \frac{1}{M}\sum_{i=0}^{M-1}\bar{R}_{in}
\end{equation}

The second term is an average by row. Therefore, every $\bar{R}_{mn}$ is subtracted by its mean along the fast time index or range bin, before storing in processed $Y_{mn}$

\begin{figure*}[]
     \centering
     \begin{subfigure}[]{0.2\textwidth}
         \centering
         \includegraphics[width=\textwidth]{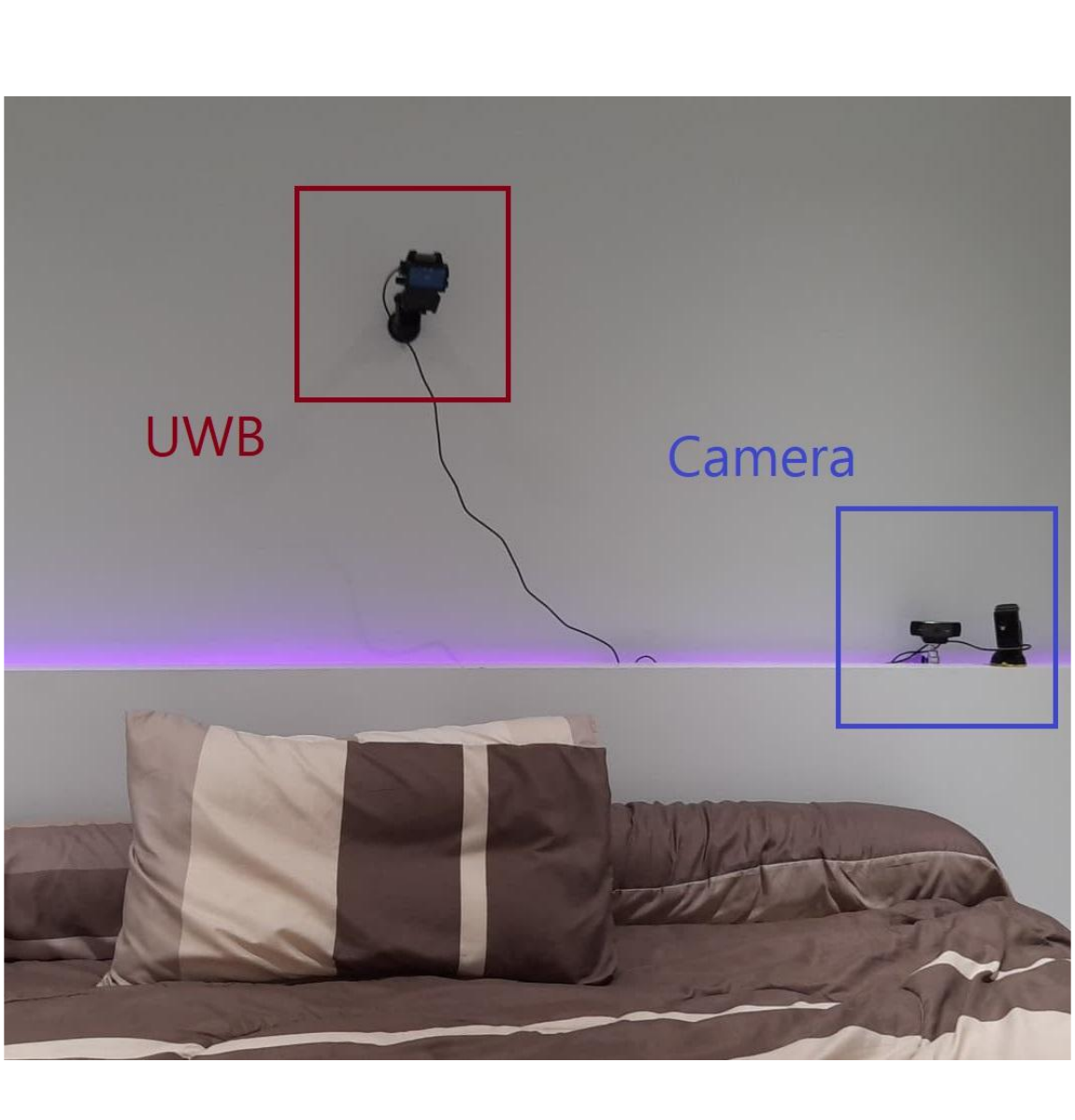}
         \caption{}
 \label{fig:bedroom}
     \end{subfigure}
     \hfill
     \begin{subfigure}[]{0.38\textwidth}
         \centering
         \includegraphics[width=\textwidth]{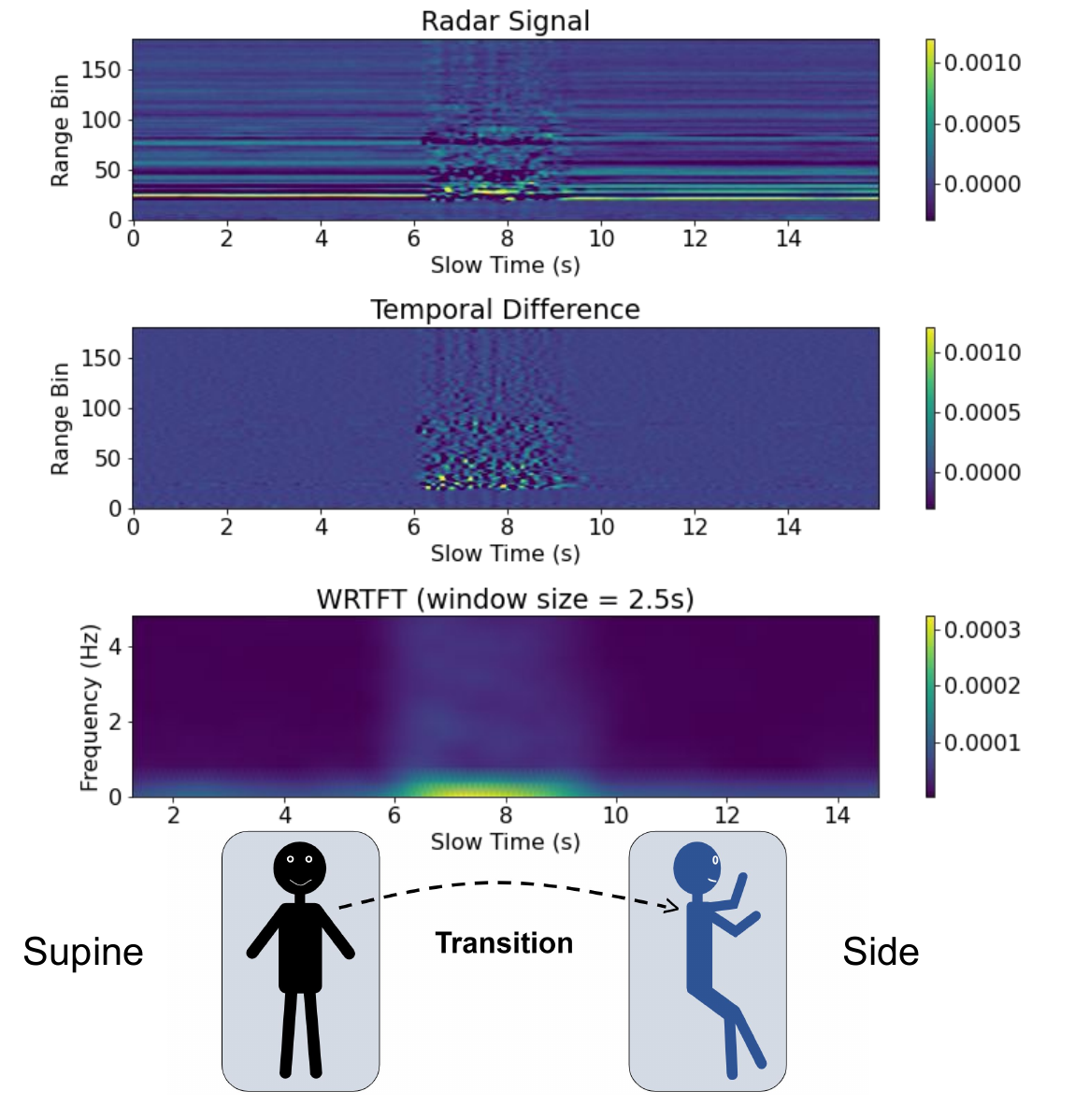}
         \caption{}
         \label{fig:data1}
     \end{subfigure}
     \hfill
     \begin{subfigure}[]{0.38\textwidth}
         \centering
         \includegraphics[width=\textwidth]{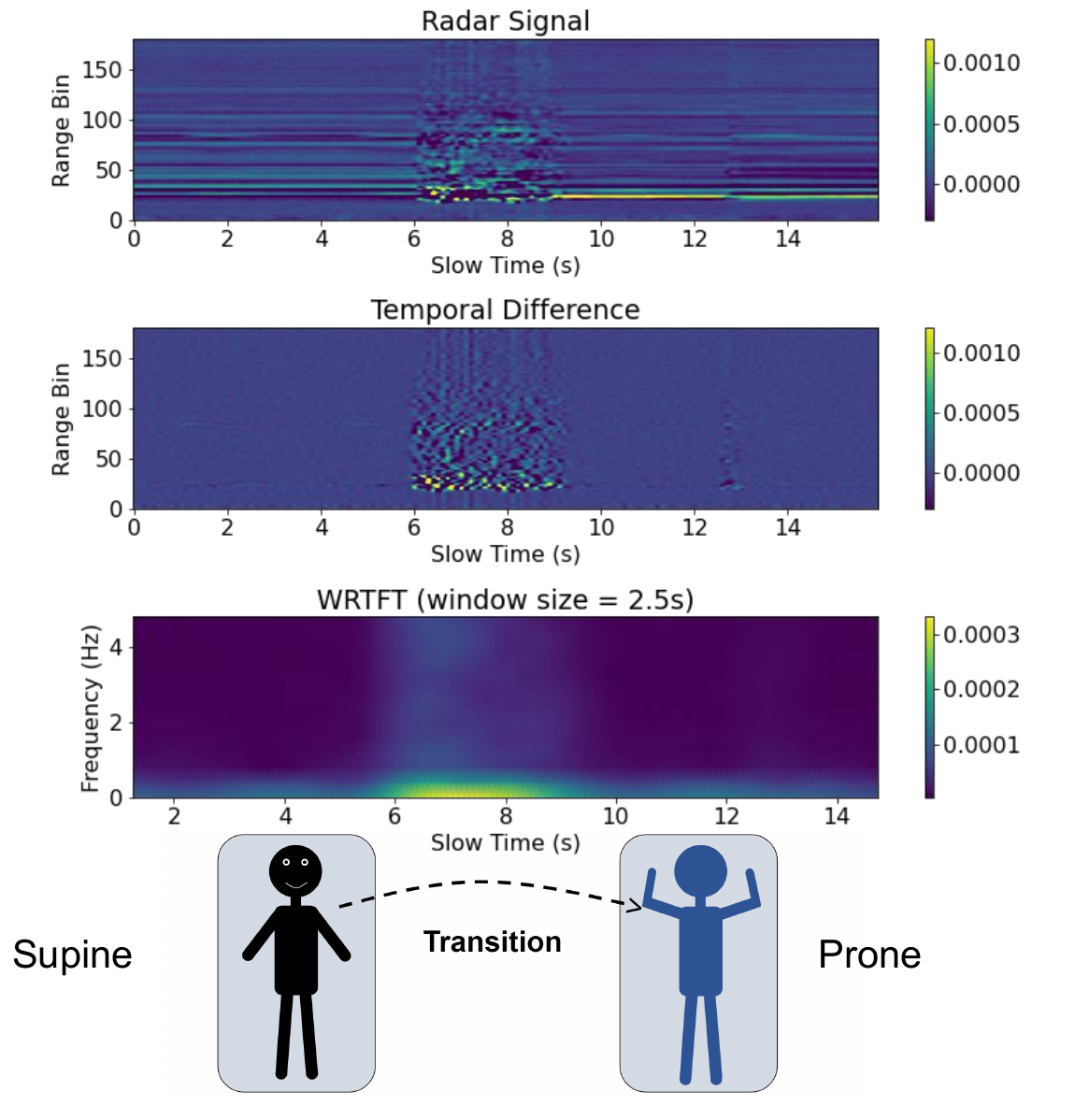}
         \caption{}
         \label{fig:data2}
     \end{subfigure}
        \caption{(a) Experimental bedroom-like environment with UWB radar devices placed above the headboard. The camera was also placed for recording as ground truth. (b) signals and features of SUSI (c) signals and features of SUPR}
        \label{fig:data}
\end{figure*}

\subsection{Feature Extraction}
In this paper, features in time domain and frequency domain are proposed to be implemented in our classification task.
\subsubsection{Time Difference (TD)} This method is based on an assumption that every obstacle and object in a bedroom-like environment are static or not changing its shape over the time. Therefore, the differentiation along the slow time axis $Y^{d}$ can extract only the information of a moving human, as shown in \autoref{fig:data1} and \autoref{fig:data2}. Most non-transitioning parts are suppressed by this finite differentiation.
\begin{equation}\label{eq:5}
Y^{d}_{mn} = Y_{m,n+1} - Y_{mn}
\end{equation}

Where $Y^{d}_{mn}$ is the entry in $m$-th row and $n$-th column of matrix $Y^{d}$. Every $Y^{d}_{mn}$ is a difference between $Y_{m,n+1}$ and $Y_{mn}$, which represents the pulses from range bin $m$ with different slow time.

\subsubsection{Range Selection} For our device, the step size between range is 5.14 cm. 40 range bins containing information in range of about 2 m and covering all parts of the human body are chosen. The selection algorithm is done by finding a position of cropping window such that maximized the summation of the slow time difference energy in the window size of $(F-I+1) \times N$, as shown in \autoref{eq:range_optimization}.

\begin{equation}\label{eq:range_optimization}
\begin{aligned}
& \underset{I, F}{\text{maximize}}
& & \sum_{m=I}^{F}\sum_{n=0}^{N-1}(Y^{d}_{mn})^{2} \\
& \text{participant to} & &  F-I+1 = 40
\end{aligned}
\end{equation}

Where $I$ and $F$ are the initial range bin and final range bin of the cropping window. A cropped signal is then stored in the $(F-I+1) \times N $ matrix.

\subsubsection{Weighted Range-Time-Frequency Transform (WRTFT)}
WRTFT was proposed in 2018 \cite{ding2018non}, to combine spectrograms from all range bins. After the transformation is complete, the output image contain information from range, time, and frequency features of the human motions. WRTFT is performed according to the following steps;

\paragraph{Short Time Fourier Transform (STFT)} 
STFT is performed by segmenting time series and employing Fourier transform (FT) on each segment. The obtained result is referred to time-frequency representation (TFR), and each range bin hold its own TFR. The result is stored in three-dimensional array $F$.

\begin{equation}\label{eq:7}
F_{mkn} = \sum_{p=0}^{N-1}Y_{mp}\omega(n-p)e^{-j2\pi pk/N}, m \in [I, F]
\end{equation}
where $k$ is frequency domain index and $\omega(p)$ is a window function.

\paragraph{Weighted Average}
All spectrograms from all range bins are weighted-averaged by energy signal of each range bin and stored in matrix $W$;
\begin{equation}\label{eq:8}
W_{kn} = \sum_{m=I}^{F}\sigma_{m}F_{mkn}
\end{equation}

\begin{equation}\label{eq:9}
\sigma_{m} = E_m/\sum_{m=I}^{F}E_{m}
\end{equation}

\begin{equation}\label{eq:10}
E_{m} = \sum_{n=0}^{N-1}Y_{mn}^{2}
\end{equation}

where $\sigma_{m}$ is the coefficient weighting a spectrogram and is calculated by the proportion of energy in a range bin.

TD and WRTFT of SUSI and SUPR are shown in \autoref{fig:data1} and \autoref{fig:data2}. Both time and frequency features are unlikely to contribute in visually distinguishing between the classes.

\begin{figure*}[]
    \centering
     \begin{subfigure}[]{0.18\textwidth}
         \centering
         \includegraphics[width=0.9\textwidth]{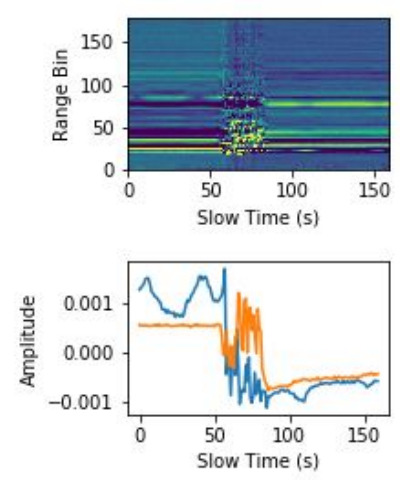}
         \caption{}
         \label{fig:aug_noaug}
     \end{subfigure}
     \hfill
     \begin{subfigure}[]{0.18\textwidth}
         \centering
         \includegraphics[width=0.9\textwidth]{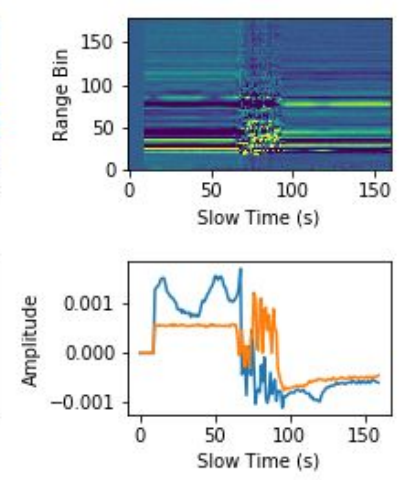}
         \caption{}
         \label{fig:aug_ts}
     \end{subfigure}
     \hfill
     \begin{subfigure}[]{0.18\textwidth}
         \centering
         \includegraphics[width=0.9\textwidth]{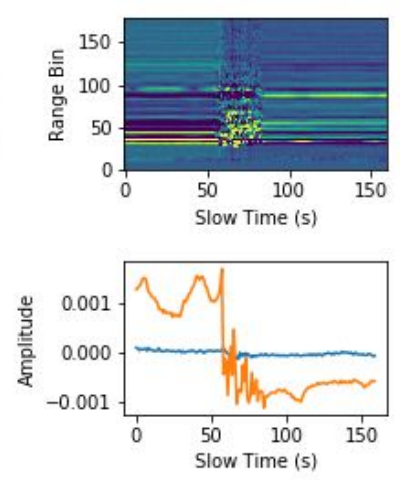}
         \caption{}
         \label{fig:aug_rs}
     \end{subfigure}
     \hfill
     \begin{subfigure}[]{0.18\textwidth}
         \centering
         \includegraphics[width=0.9\textwidth]{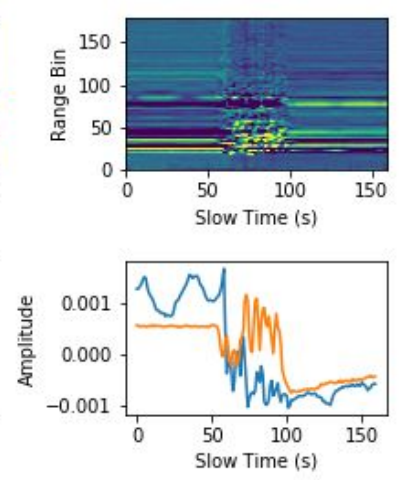}
         \caption{}
         \label{fig:aug_tw}
     \end{subfigure}
     \hfill
     \begin{subfigure}[]{0.18\textwidth}
         \centering
         \includegraphics[width=0.9\textwidth]{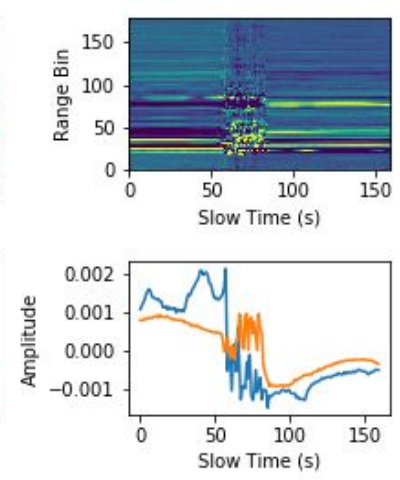}
         \caption{}
         \label{fig:aug_mw}
     \end{subfigure}
\caption{Four different data augmentation methods were applied to prevent the occurrence of overfitting. Upper low displays raw radar signals and lower row displays the two signals of different range bins as the result of (a) without the augmentation in comparison to four augmentation methods: (b) Time Shift , (c) Range Shift , (d) Time-Warping , and (e) Magnitude-Warping.}
        \label{fig:augmentation}
\end{figure*}

\subsection{Data Augmentation}
Deep learning normally requires a large training dataset to prevent the occurrence of overfitting. However, for human bio-signals, data are commonly small due to the limited human participants and resources. In order to solve this problem, Time Shift (TS) and Range Shift (RS) were applied. In addition, Time-Warping (TW) and Magnitude-Warping (MW), which were proposed in 2017 for wearable sensor data \cite{um2017data}, were also used. The results from data augmentation methods are shown in \autoref{fig:augmentation}. All possible combinations from these four augmentation methods were applied to the training data.

\paragraph{Time Shift (TS)} Approximately 2-3 s segments, containing SPT information, are in a long interval of slow time in one sample. The shift on slow time index can possibly assist the model to learn a SPT section in a different position. The parameters for TS include [-10, -5, 5, 10], where positive and negative values indicate right and left shifts of slow time index respectively, and an extension section is padded with zeros.
\paragraph{Range Shift (RS)}
There are different range bins in many samples which represent the respective part of the human body. RS is implied to be capable of aiding in learning the variability of human's range position. The parameters for RS included [2, 4], where positive values indicate the up shift of range bin and extension section is also padded with zeros.
\paragraph{Time-Warping (TW)} TW is done by smoothing and randomly distorting the intervals between slow time indices in each sample to change the temporal location of the slow time indices. Cubic spline interpolation is used to impute the missing temporal location values by their neighbors, which are the original value with different temporal location. The variance parameter used in this work is 0.4, representing the variance of interval distortion.
\paragraph{Magnitude-Warping (MW)} MW create a random smooth curve varying around one and each sample is multiplied with this curve in order to randomly change the amplitude along slow time positions. The variance parameter used in this work is 0.4, representing the variation around one in a random curve.

\begin{figure*}[]
\centering
\includegraphics[width=0.9\textwidth]{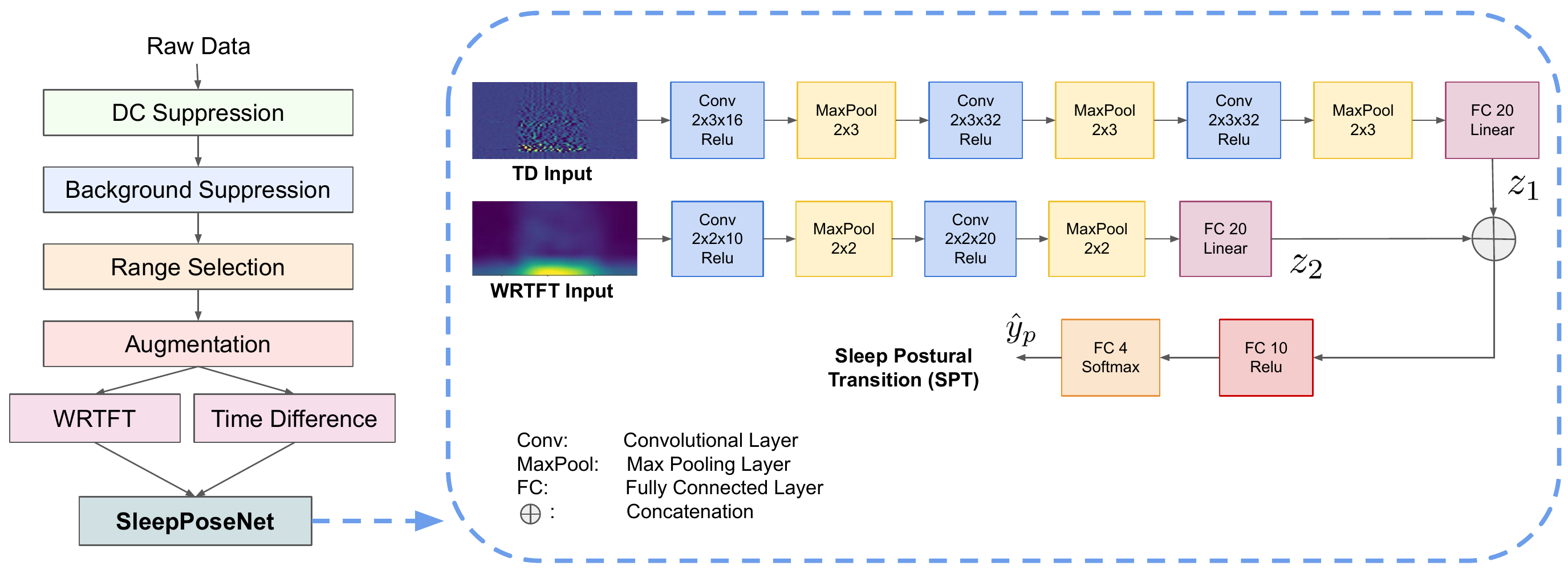}
\caption{{\color{blue}Flow chart shows the overview of the complete procedure including data processing, range selection, data augmentation, and feature extraction, before flowing through our proposed model, SPN. The model architecture is also shown in blue-dash-line} window.}
\label{fig:sleepposenet}
\vspace{-0mm}
\end{figure*}

\subsection{SleepPoseNet}
The overview of all processes from the data processing to prediction are illustrated as the flow chart in \autoref{fig:sleepposenet}.
In order to extract significant features from two views of data and perform two tasks simultaneously, this study proposed a method, entitled SleepPoseNet or SPN, composing of a combination of Deep Convolutional Neural Networks (DCNN) and Multi-View Learning (MVL).

DCNN is a class of neural networks, consisting of convolutional filters, which perform arithmetic operations to find relationship of a data point, and its neighbors, to extract some important features from all data points in a sample. DCNN works especially well in image recognition tasks. Different DCNN models may include 1D convolutional filters, which are used for time series classification and prediction.
The data collected in this study are multivariate time series, which can be represented by 2D array similarly to an image. Therefore, DCNN with 2D convolutional filters for the classification tasks is deemed appropriate.


MVL combines and utilizes the information from different views to improve the generalization performance. For instance, video and sound together may contribute to higher classification accuracy than only a single view. In this paper, we combined the information from time and frequency domains of the radar signal to gain higher SPT classification accuracy. 





{\color{blue}
The architecture of SPN is illustrated in \autoref{fig:sleepposenet}. Here, the attempt to fuse the information from both time and frequency domains to increase the classification accuracy was executed. Features from TD and WRTFT were separately extracted by early convolutional layers of the neural networks. Consequently, the extracted features were fused by concatenation and passed through the late dense layers. The parameters of model are displayed in \autoref{fig:sleepposenet}. The filter sizes were set as 2x3 for TD and 2x2 for WRTFT, and the max pooling layers were used to reduce the dimension, preventing the overfitting with same filter sizes. Cross entropy loss in Equation \ref{eq:posture_loss} was minimized to achieve higher classification accuracy.

\begin{equation}\label{eq:posture_loss}
J_{post}(\theta) = -\frac{1}{M}\sum^{M}_{i=1}{y}_{p}^{(i)}\log \hat{y}_{p}^{(i)}(\theta)
\end{equation}
}

\subsection{Experiments}
{\color{blue}
In this section, comparison methods were employed to evaluate SPN, partitioning into four experiments. The first three experiments were evaluated with Dataset I, while the last experiment was performed on Dataset II. Experiment I and Experiment II investigated the SPN performance in comparison to other models among different distributions between training, validation, and test data. Experiment III examined the range selection process of the number of range bins, also called window size (WS), which affects the performance of SPN. Experiment IV investigated more naturalistic situations where other activities (BG class) and a moving object were included. Here, various methods based on WRTFT \cite{ding2018non, chen2019non} were chosen to compare to our MVL method with a variety of configurations.
}
\begin{figure}[]
     \centering
     \begin{subfigure}[]{0.22\textwidth}
         \centering
         \includegraphics[width=\textwidth]{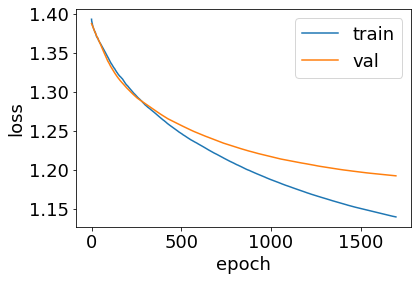}
         \caption{WRTFT-CNN}
         \label{fig:loss_exp1_wrtft}
     \end{subfigure}
     \hfill
     \begin{subfigure}[]{0.22\textwidth}
         \centering
         \includegraphics[width=\textwidth]{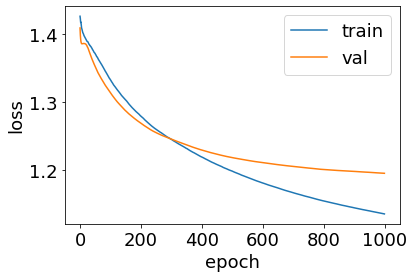}
         \caption{TD-CNN}
         \label{fig:loss_exp1_td}
     \end{subfigure}
     \hfill
     \begin{subfigure}[]{0.22\textwidth}
         \centering
         \includegraphics[width=\textwidth]{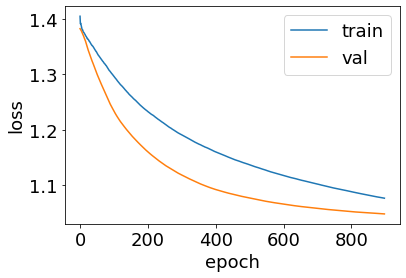}
         \caption{SPN}
         \label{fig:loss_exp1_multiview}
     \end{subfigure}
     \hfill
     \begin{subfigure}[]{0.22\textwidth}
         \centering
         \includegraphics[width=\textwidth]{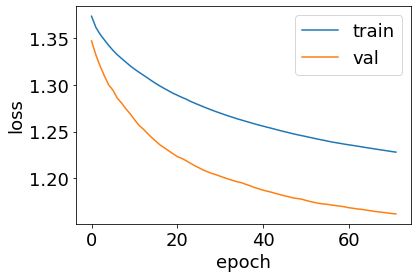}
         \caption{AUG-WRTFT-CNN}
         \label{fig:loss_exp1_aug_wrtft}
     \end{subfigure}     
     \hfill
     \begin{subfigure}[]{0.22\textwidth}
         \centering
         \includegraphics[width=\textwidth]{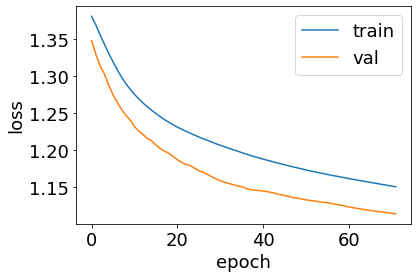}
         \caption{AUG-TD-CNN}
         \label{fig:loss_exp1_aug_td}
     \end{subfigure}
     \hfill
     \begin{subfigure}[]{0.22\textwidth}
         \centering
         \includegraphics[width=\textwidth]{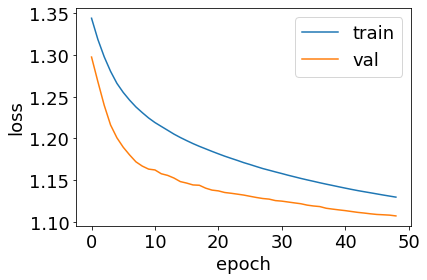}
         \caption{AUG-SPN}
         \label{fig:loss_exp1_aug_multiview}
     \end{subfigure}   
        
        \caption{
        {\color{blue}Training and validation losses from all methods as described in Experiment I: Unseen Test Participants.}
        }
        \label{fig:loss_exp}        
\end{figure}

\begin{figure*}
     \centering
     \begin{subfigure}[]{0.3\textwidth}
         \centering
         \includegraphics[width=\textwidth]{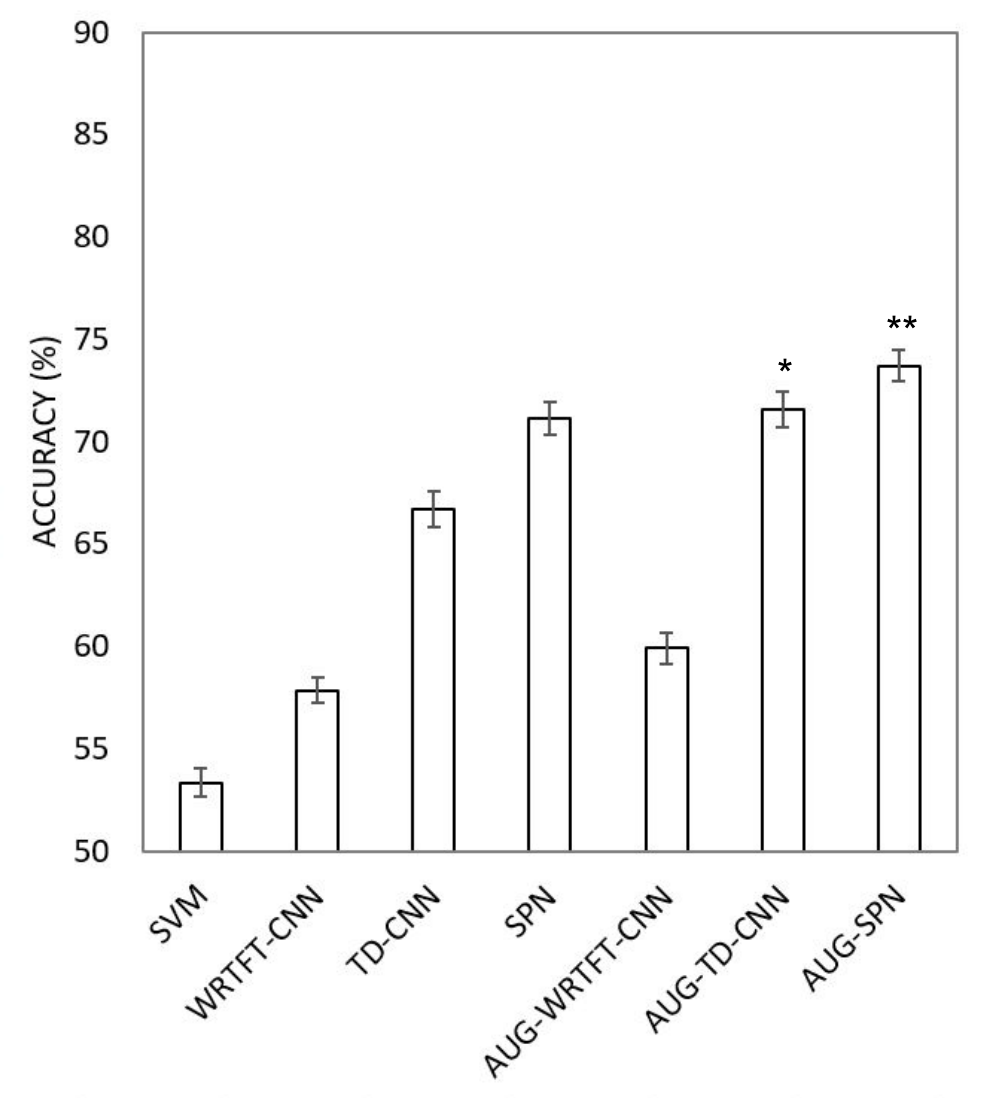}
         \caption{}
         \label{fig:exp1_acc}
     \end{subfigure}
     \centering
     \hfill
     \begin{subfigure}[]{0.3\textwidth}
         \centering
         \includegraphics[width=\textwidth]{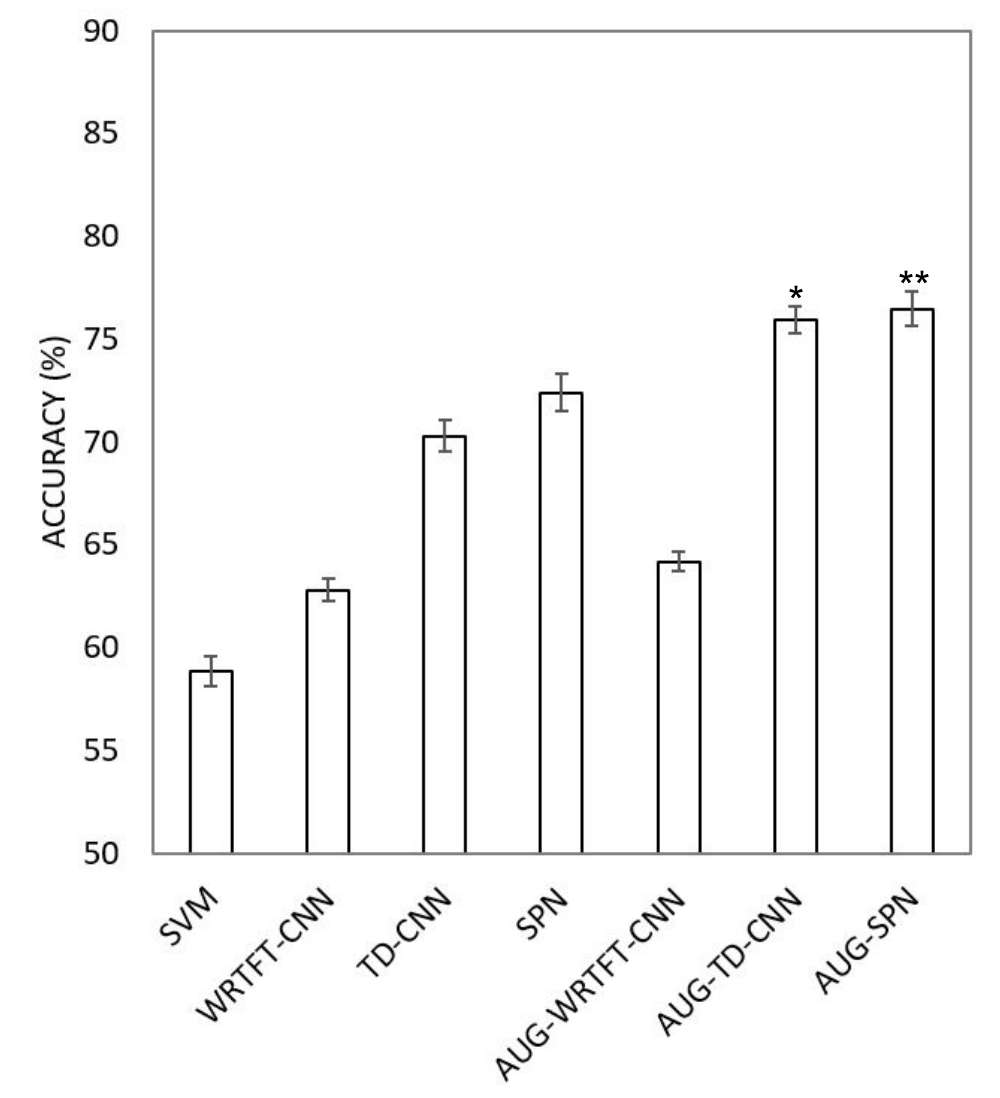}
         \caption{}
         \label{fig:exp2_acc}
     \end{subfigure}
    \centering
    \hfill
     \begin{subfigure}[]{0.3\textwidth}
         \centering
         \includegraphics[width=\textwidth]{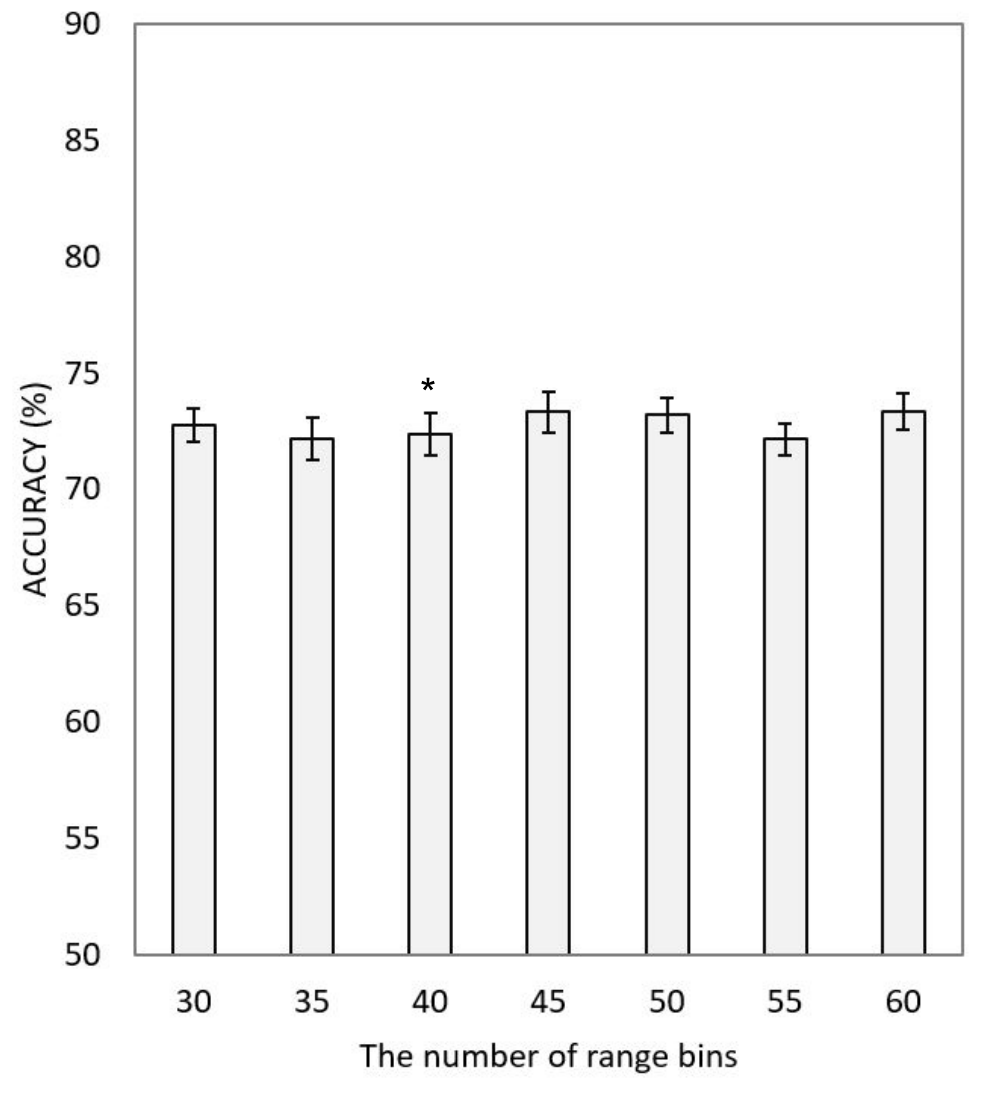}
         \caption{}
         \label{fig:exp3_acc}
     \end{subfigure}
    
  \caption{Comparison of the accuracy rates in percentage with standard errors (SE) from Experiment I and Experiment II. (a) Unseen test participant (b) Seen test participant and (c) Experiment III}
\label{fig:result_acc_exp4}
\end{figure*}

\begin{figure}
     \centering
     \begin{subfigure}[]{0.4\textwidth}
         \centering
         \includegraphics[width=\textwidth]{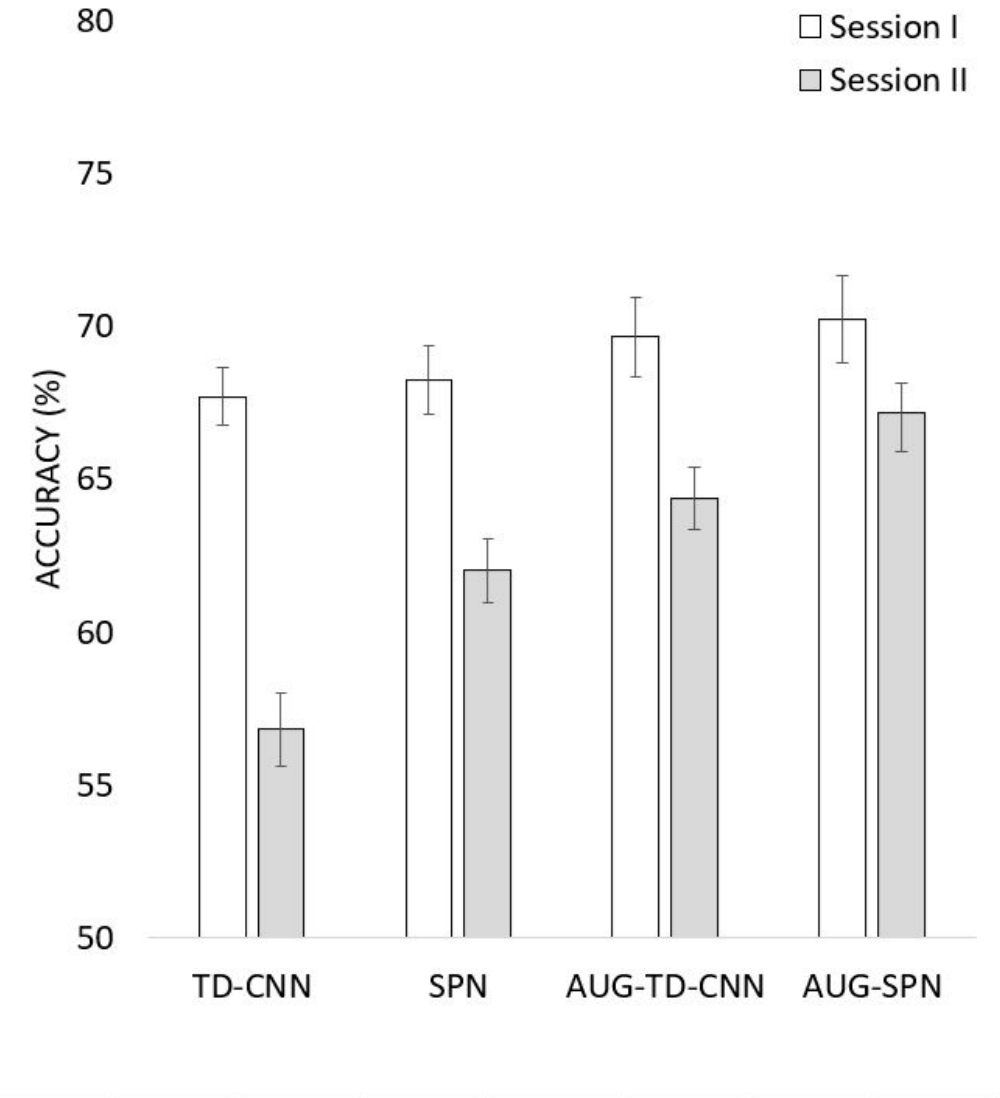}
         \caption{}
         \label{fig:exp4_unseen_acc}
     \end{subfigure}
     \centering
     \hfill
     \begin{subfigure}[]{0.4\textwidth}
         \centering
         \includegraphics[width=\textwidth]{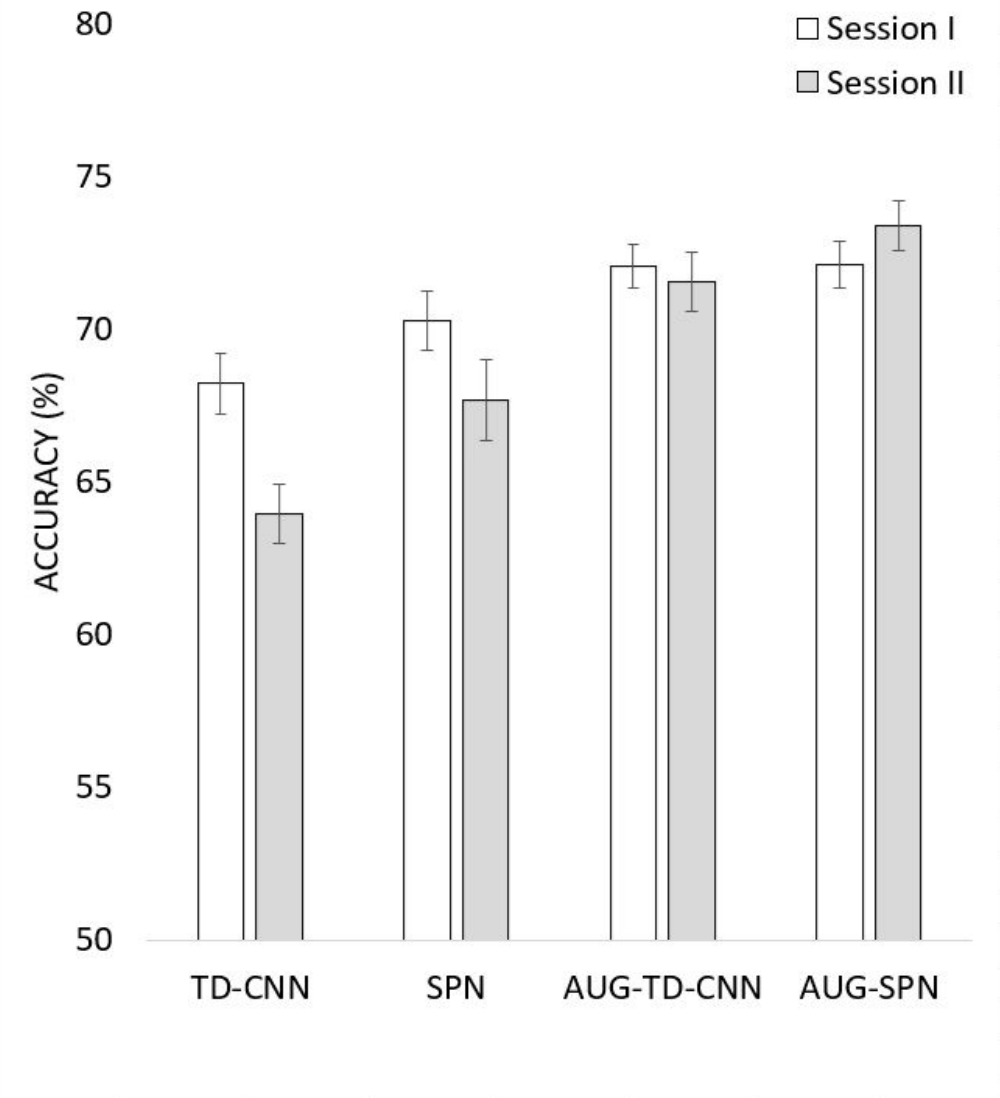}
         \caption{}
         \label{fig:exp4_seen_acc}
     \end{subfigure}

  \caption{Comparison of the accuracy rates in percentage with standard errors (SE) from Experiment IV, (a) Unseen test participant, and (b) Seen test participant}
\label{fig:result_acc}
\end{figure}

\begin{figure}[]
     \centering
     \begin{subfigure}[]{0.24\textwidth}
         \centering
         \includegraphics[width=\textwidth]{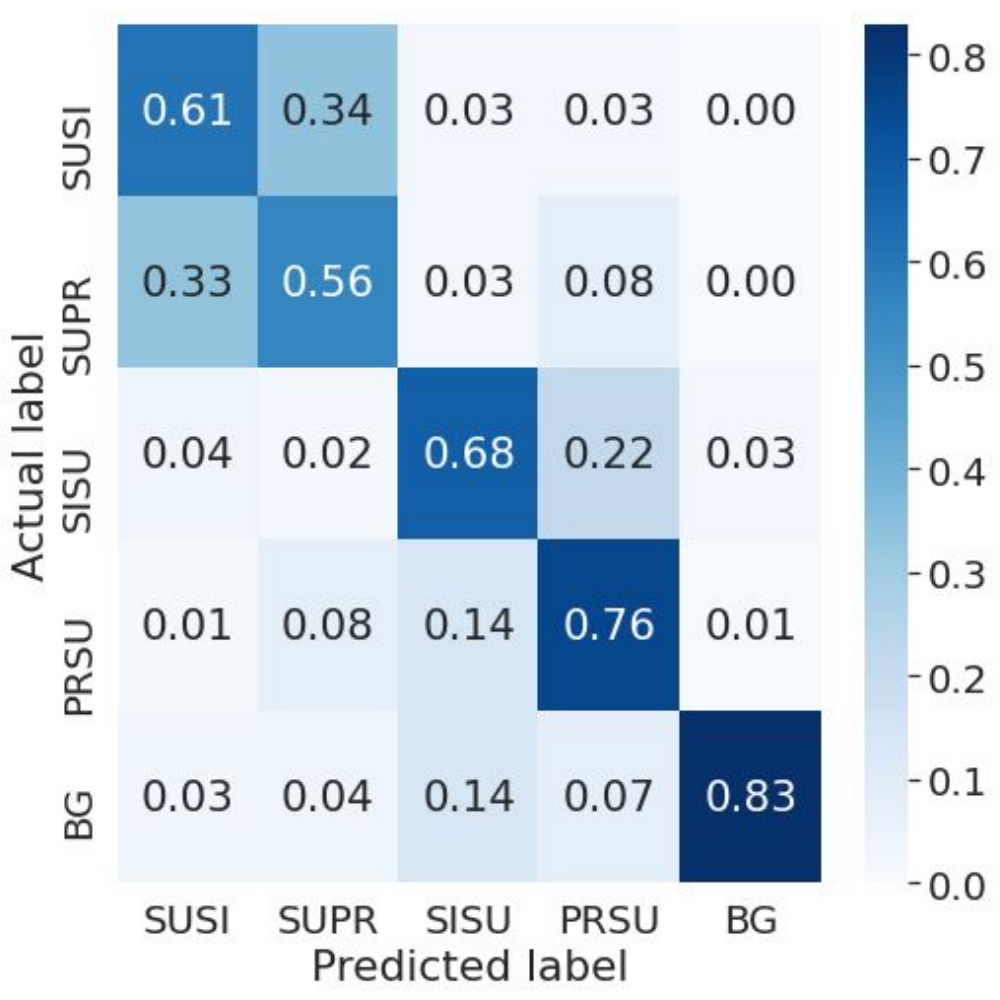}
         \caption{}
         \label{fig:unseen_cm1}
     \end{subfigure}
     \hfill
     \begin{subfigure}[]{0.24\textwidth}
         \centering
         \includegraphics[width=\textwidth]{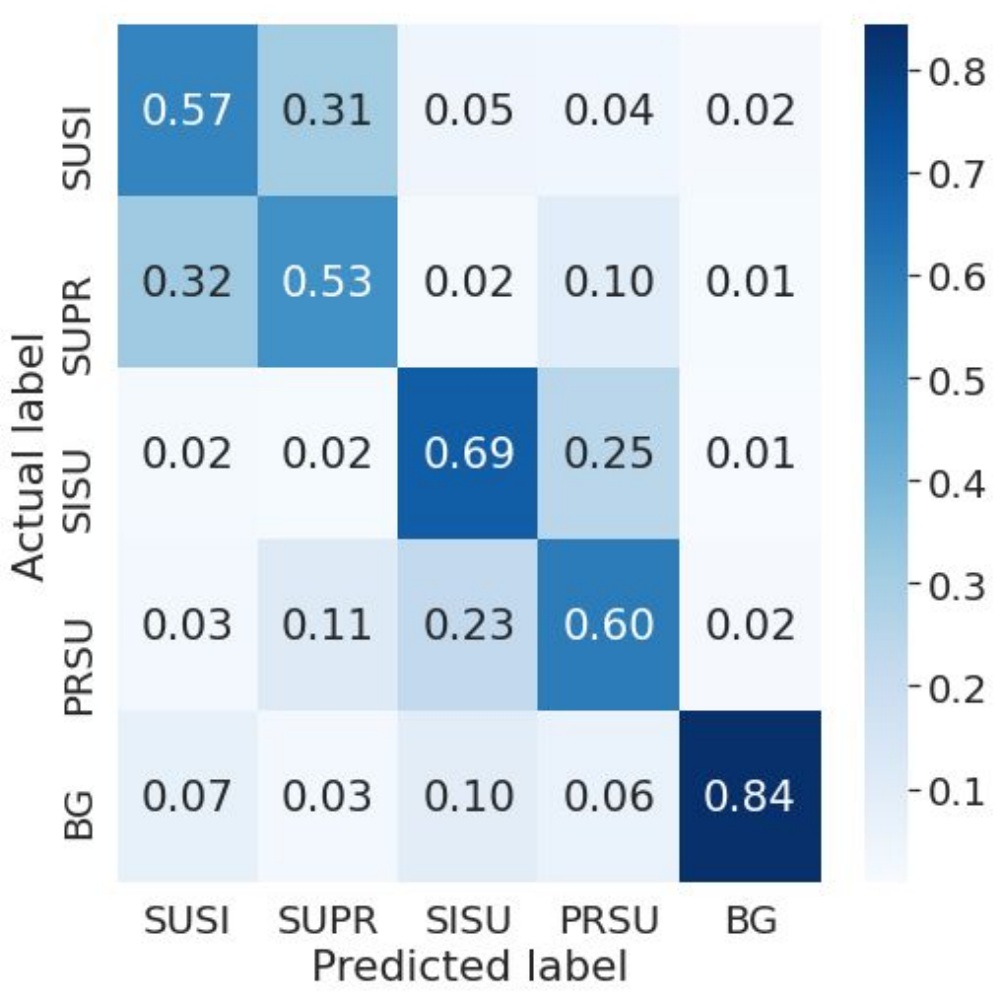}
         \caption{}
         \label{fig:unseen_cm2}
     \end{subfigure}
     \hfill
     \begin{subfigure}[]{0.24\textwidth}
         \centering
         \includegraphics[width=\textwidth]{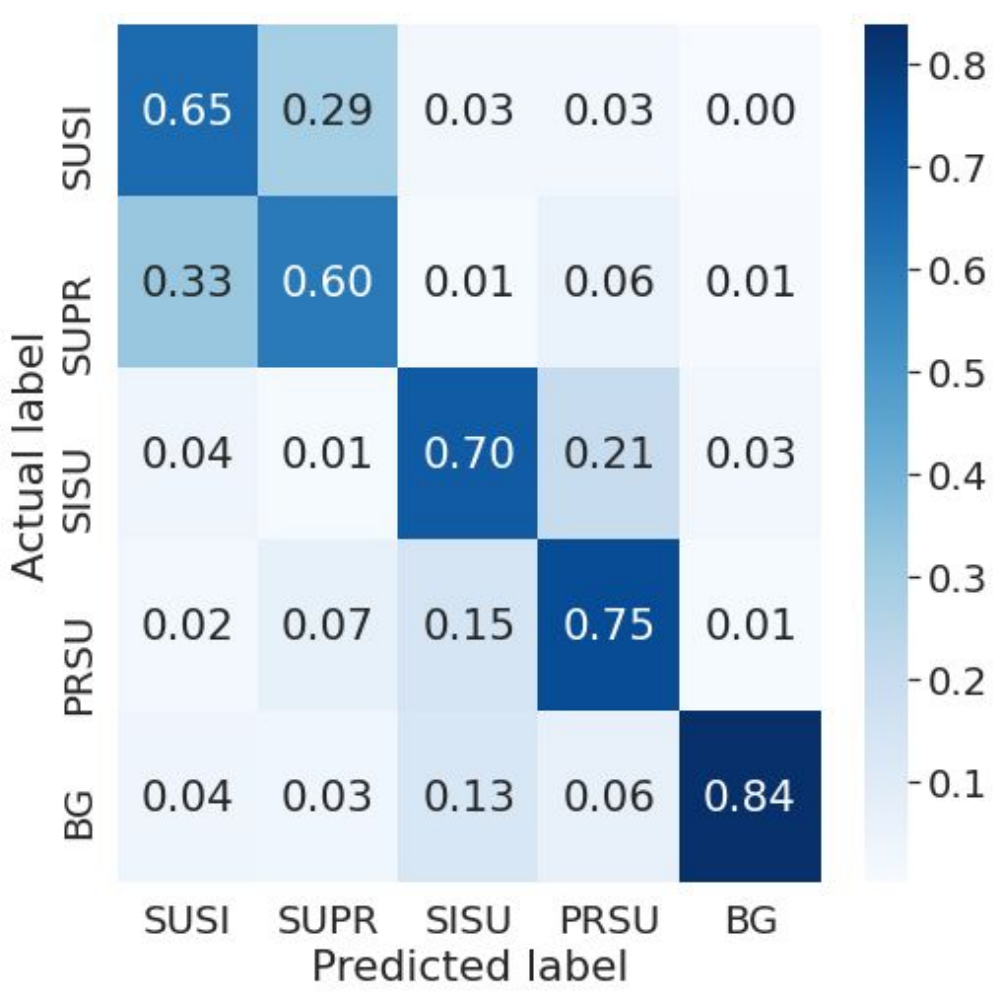}
         \caption{}
         \label{fig:seen_cm1}
     \end{subfigure}
     \hfill
     \begin{subfigure}[]{0.24\textwidth}
         \centering
         \includegraphics[width=\textwidth]{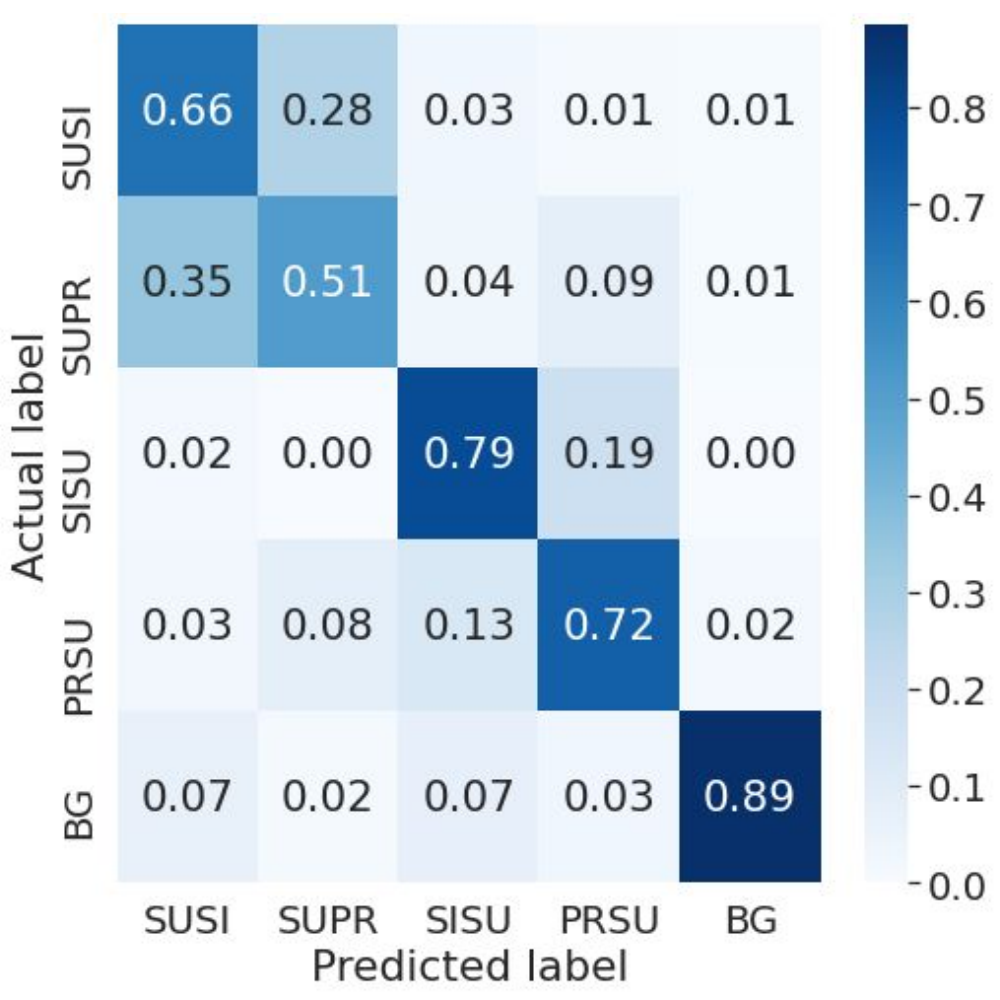}
         \caption{}
         \label{fig:seen_cm2}
     \end{subfigure}     
     
        \caption{Confusion matrices of AUG-SPN in four scenarios in Experiment IV. (a) Unseen test participant and Session I and (b) Unseen test participant and Session II (c) Seen test participant and Session I (d) Seen test participant and Session II}
        \label{fig:cm_exp4}        
\end{figure}

\textit{Model Evaluation:} {\color{blue} 
Four methods, including SPN, were selected for evaluating the performance of SPN. Firstly, SVM with Radial Basis Function (RBF) kernel was used along with the first 20 principal components of WRTFT, extracted using PCA. Grid-search cross-validation was applied to find the best parameters of SVM. The search sets of C and gamma were $\{0.01,0.1,1,10,100\}$ and $\{0.01,0.1,1,10,100\}$, respectively. WRTFT-CNN was employed, composing of a convolutional layer with 10 filters and a max pooling layer as well as a convolutional layer with 20 filters and a max pooling layer, respectively. All filters were set with the size of 2 by 2, ending with two dense layers of 20 and 10 hidden units, followed by rectified linear unit (ReLU) activation, and 4 output units, followed by softmax activation. TD-CNN was composed of three convolutional layers with 16, 32, and 32 filters and three max pooling layers with the size of 2 by 3. In other word, WRTFT-CNN's and TD-CNN's parameters were the same as the bottom and top part of SPN respectively. All methods were also trained with augmented training data. Hence, the augmented adaptations of aforementioned methods were named AUG-WRTFT-CNN, AUG-TD-CNN, and AUG-SPN. All schemes were run for five times with different random states. To regularize the models, spatial dropout and dropout with the probability of 0.3 and 0.5 were employed on every convolutional layer and every hidden layer, which were placed before output layers. All models were trained by ADAM optimization with a constant learning rate of 0.0001. A small batch size of 16 was used along with early stopping. The averaged accuracy of all methods from all folds were then compared and statistically tested using one-way repeated measure ANOVA.}

\textit{Experiment I:} {\color{blue} We first investigated the performance of all models in predicting the data from the unseen test participant. 26 participants were divided randomly for 10 times into 18, 4, and 4 participants to be used as training, validation, and test sets, respectively.}

\textit{Experiment II:} {\color{blue}
To examine the performance of all models with the prior knowledge of participants’ data distribution before the test data were predicted, in this experiment, we divided the training, validation, and test set using five-fold cross-validation. Each set consisted of the data from all participants.}

\textit{Experiment III:} {\color{blue}
To determine the most appropriate Window Size (WS) for range bin selection, the method with the most reliable result was examined with WS adjustment. The WS in this experiment were 30, 35, 40, 45, 50, 55, and 60. The experimental setting was identical to Experiment II.}

\textit{Experiment IV:}{\color{blue} Dataset II composing of recordings in scenarios with dynamic setting were investigated in Experiment IV. The method with preeminent performance from Experiment I and II was used in this experiment. The training and validation data composed of two sessions. Session I was performed in a static environment. Whereas in Session II, a swinging fan was placed to simulate a dynamic environment in the real-world. To evaluate an effect of the moving object, test data was divided into two sessions. An effect of test participant distribution (seen or unseen test participants) was also examined in this experiment. Leave-one-participant-out cross-validation was deployed in unseen test participant setting and six-fold cross validation was applied in seen test participant setting with maintaining data from all participants in training, validation, and test set.}

\section{Results}

{\color{blue}
All training and validation losses over the epoch in Experiment I are shown in \autoref{fig:loss_exp}. The models, with minimum validation losses, were selected. Augmented training data were shown to cause training losses to be significantly higher than validation losses.        
}

\subsection{Experiment I: Unseen Test Participants}\label{subsec:unseen}

{\color{blue}
Significant differences ($F(3.672, 179.943) = 143.602, p < 0.05$) were found amongst the accuracy means of some model comparisons. As shown in \autoref{fig:exp1_acc}, SPN with data augmentation (AUG-SPN) performed with higher accuracy than the others, revealing the highest accuracy of $73.7 \pm 0.8 \% $. In contrast, the accuracy of AUG-TD-CNN was not significantly different from SPN. SPN without data augmentation achieved an accuracy of $71.6 \pm 0.9 \%$ which was significantly higher than AUG-WRTFT-CNN with data augmentation, serving as the best method based on frequency domain features of WRTFT. AUG-WRTFT-CNN yielded the accuracy of $59.9 \pm 0.7 \%$. The models in which features included only WRTFT attained the accuracy of lower than $60\%$. Data augmentation yielded the greatest effect on TD-CNN of which the accuracy rose by approximately $5\%$.

}

\subsection{Experiment II: Seen Test Participants}\label{subsec:seen}

{\color{blue}
Among the models, significant differences in accuracy means were found ($F(2.889, 69.329) = 110.500, p < 0.05$) in some implementations. As shown in \autoref{fig:exp2_acc}, SPN with data augmentation (AUG-SPN) outperformed the others with an accuracy of $76.5 \pm 0.8\%$. However, no significant difference was found among the accuracy produced by AUG-SPN and AUG-TD-CNN. The overall accuracy of the models with the seen test participants was revealed to be higher in comparison to the accuracy of the models with the unseen test participants.
}

\subsection{Experiment III: Window Size Adjustment}\label{subsec:window}

{\color{blue}
SPN was selected for WS adjustment. \autoref{fig:exp3_acc} displays the accuracy of assorted WS. All of accuracy means of the configurations were not significantly different ($F(2.620, 65.501) = 1.565, p < 0.05$). In addition, increasing the WS did not improve the accuracy. Therefore, this suggested that selecting WS of 40 initially, covering 2 m of all parts of the human body, was optimal for real-time application.}

\subsection{Experiment IV}\label{subsec:bgandnoise}
{\color{blue}
The results of this experiment were divided into two parts: unseen test participant and seen test participant. SPN, AUG-SPN, TD-CNN, and AUG-TD-CNN were candidates to be further evaluated in new scenarios. The WS of 40 was selected, similarly to \autoref{subsec:seen} and \autoref{subsec:unseen}.
}

\subsubsection{Unseen Test Participant}
{\color{blue}
There were significant differences in the accuracy means among some models ($F(4.214, 248.645) = 22.366, p < 0.05$). As shown in \autoref{fig:exp4_unseen_acc}, in general, the accuracy of each method in Session I was higher than the same method performed in Session II. In Session I, AUG-SPN achieved the highest accuracy of $70.2 \pm 1.4 \% $, followed by AUG-TD-CNN of which accuracy was $69.7 \pm 1.3 \% $. However, no significant difference was found among the four methods in Session I. In Session II, AUG-SPN also yielded the highest accuracy of $67.2 \pm 1.3 \% $. AUG-TD-CNN yielded the second highest accuracy of $64.4 \pm 1.3 \% $. Data augmentation raised the accuracy of TD-CNN and SPN significantly by approximately $8 \%$ and $5\%$, respectively. The confusion matrices of AUG-SPN in Session I and Session II are shown in \autoref{fig:unseen_cm1} and \autoref{fig:unseen_cm2}, respectively. Overall, the models in both sessions yielded the highest recall of BG and the lowest recall of SUPR. The models frequently misclassified SUSI as SUPR and SUPR as SUSI. Similarly, the models also predominantly misclassified SISU as PRSU and PRSU as SISU. The recall of the same class in Session II tended to be lower than those in Session I, with the exceptions of BG and SISU.
}

\subsubsection{Seen Test Participant}
{\color{blue}
Significant differences in the accuracy of some models were found ($F(3.645, 105.701) = 554.639, p < 0.05$). Overall, the accuracy in Session I was higher than that of Session II, with the exception of AUG-SPN as shown in \autoref{fig:exp4_seen_acc}. In Session I, AUG-TD-CNN and AUG-SPN yielded the highest accuracy of $72.1 \pm 0.7 \% $ and $72.1 \pm 0.8 \% $, respectively. TD-CNN performed with the lowest accuracy of $68.2 \pm 1.0 \% $. However, there was no significant difference between TD-CNN and SPN. In Session II, AUG-SPN yielded the highest accuracy of $73.4 \pm 0.8 \% $. This did not deem significantly different to AUG-TD-CNN with the accuracy of $71.6 \pm 1.0 \% $. \autoref{fig:seen_cm1} and \autoref{fig:seen_cm2} present the confusion matrices for AUG-SPN in Session I and Session II, respectively. In general, the models used in both sessions produced the highest recall of BG and the lowest recall of SUPR. Similar to the results of unseen test participant, SUSI was generally misclassified as SUPR while SUPR was commonly misclassified as SUSI. Moreover, misclassification between SISU and PRSU had the same tendency as between SUSI and SUPR like the results in unseen test participant. The recall of the same class in Session II was demonstrated to be higher than those in Session I, with the exception of SUPR and PRSU.
}

\section{Discussion}

{\color{blue}
The attractiveness of UWB radar system for non-contact human activity detection has led this study to adapt the radar-based signals for the recognition of sleep postures. Using a novel designed SleepPoseNet (SPN), we evaluated the performance of different WRTFT-based models. The possibility of developing and further implementing the UWB sensors for monitoring patients with sleep disorders was observed.

Here in this study, the Xethru UWB radar, an off-the-shelf device, was selected for usage due to its affordable cost, user-friendly attribute, and easy-to-use element suitable as a smart-home device. However, the device has a single Rx/Tx antenna that may induce some limitations in receiving information. For instance, it cannot distinguish objects with the same range but different angles. It is imperative to improve its efficiency by gathering information from more individuals. This work serves as the instigation for further development of the monitoring aspect of the device. To commence, data from 38 participants were first collected and digital signal processing with deep learning approach was performed for the task analysis.

Previous studies have been utilizing WRTFT with DCNN for human activity classification by using UWB radar signals \cite{ding2018non, chen2019non}. Using their works as the foundation, we attempted to further improve the classification performance by incorporating higher complexity learning methods to our model, i.e., MVL as shown in \autoref{fig:sleepposenet}. \autoref{fig:exp1_acc} demonstrates the experimental results, showing that the proposed SPN was able to achieve 73.7\% accuracy, which outperformed the previous method with an improvement of 13.8\%. The amount of acquired data generally imposes an issue for the accuracy of the performance. Here, we applied the data augmentation methods for time series, resulting in a higher quality performance.

Intriguingly, SPN was demonstrated to perform better than the state-of-the-art model. From Experiment I and II, giving some examples from every participant to the model was shown to improve the performance better than leave-participant out approach. Thus, to perform with much higher accuracy, a registration system which requires some information before execution can be beneficial in practicality. Additionally, the result of Experiment III demonstrated that using the WS of 40 increased the accuracy of the models. Nonetheless, increasing the WS does not improve performance. With more data collected, the trade-off should be examined and the association of more than 4 SPTs should be considered in future studies for more practical usage in a real-world environment. Moreover, the experiment IV demonstrated that other activities or background (BG) slightly reduced classification performance as shown in \autoref{fig:cm_exp4} and moving objects can dramatically degrade classification performance in many scenarios as shown in \autoref{fig:result_acc_exp4}. \autoref{fig:cm_exp4} also inferred the ambiguity between the pairs of SUSI and SUPR and SISU and PRSU. Misclassified samples are mostly in their pair because activities in each pair share the same moving direction. Unlike human activities in previous works \cite{ding2018non, chen2019non} of which differences between classes can be visually observed through WRTFT images, SPTs are very similar as shown in \autoref{fig:data} and improvement of SPTs classification accuracy is still challenging. The future direction will probably be using two antennas which can separate the multiple objects with the same ranges; therefore, it may improve classification accuracy.

The authors wished to set this work as an exemplar with an impact on UWB radar applications and escalate the benefits of this technology for future usage. With the increase in development of the off-the-shelf UWB radar systems, the installation procedure of the systems for monitoring purpose in buildings can reduce the difficulty of utilization. This, therefore, eases the usage transitioning to increase the quality of the classification model’s performance by supplementing other factors such as heart rate or respiratory rate. Moreover, according to the findings from the previous works, extracting both spatial and time domain features from the human bio-signals is beneficial in deep learning approaches and can be applied in various tasks \cite{ditthapron2019universal, wilaiprasitporn2019affective}. Using transfer learning is another interesting addition which may aid the model to continue learning when equipped with more devices \cite{banluesombatkul2020metasleeplearner}.
Additionally, from previous studies of our team members \cite{pirbhulal2018heartbeats,wu2018quantitative}, we employed physiological signals from contact sensors for biometric application and stress monitoring. These devices are deemed compatible with non-contact body sensor, i.e., UWB radar. Furthermore, the non-contact equipment has been shown to benefit the security issue of assessing the biometric signals from wireless and implantable devices in patients \cite{8432063}. These formulate an excellent adaptation for model improvement in UWB radar applications in the future.

}
\section{Conclusion}
{\color {blue}
This study proposed SleepPoseNet (SPN), a novel core architecture featuring Multi-View (MVL) deep learning approaches, to classify different human sleep postures utilizing the signals from single ultra-wideband (UWB) antenna. The classification results of the four sleep postural transitions (SPTs) demonstrated promising recognition. Moreover, we incorporated time series data augmentation techniques to prevent an overfitting problem during the training session of SPN. This prevention significantly enhanced the classification performance. In summary, SPN will contribute as a pioneer deep learning architecture for various UWB-based applications including human sleep monitoring.
}

\vspace{-2.2mm}\bibliographystyle{IEEEtran}

\bibliography{ref} 

\end{document}